\documentclass[aps, onecolumn, superscriptaddress]{revtex4-2}

\usepackage{amsmath,amssymb}
\usepackage{bm}
\usepackage{graphicx,epsfig}
\usepackage{lineno}
\usepackage[dvipsnames]{xcolor}
      
\begin{document}
\title{Point torque representations of ciliary flows}

\author{Siluvai Antony Selvan}
\affiliation{School of Mathematics and Statistics, The University of Melbourne, Parkville, Victoria 3010, Australia.}
\affiliation{Department of Mathematics, The University of Manchester, Oxford Road, Manchester M13 9PL, UK.}

\author{Peter W. Duck}
\affiliation{Department of Mathematics, The University of Manchester, Oxford Road, Manchester M13 9PL, UK.}

\author{Draga Pihler-Puzovi\'{c}}
\email{draga.pihler-puzovic@manchester.ac.uk}
\affiliation{Department of Physics and Astronomy and Manchester Centre for Nonlinear Dynamics, The University of Manchester, Oxford Road, Manchester M13 9PL, UK.}

\author{Douglas R. Brumley}
\email{d.brumley@unimelb.edu.au}
\affiliation{School of Mathematics and Statistics, The University of Melbourne, Parkville, Victoria 3010, Australia.}

\date{\today}

\begin{abstract}
Ciliary flows are generated by a vast array of eukaryotic organisms, from unicellular algae to mammals, and occur in a range of different geometrical configurations. We employ a point torque -- or `rotlet' -- model to capture the time-averaged ciliary flow above a planar rigid wall. We demonstrate the advantages (i.e. accuracy and computational efficiency) of using this, arguably simpler approach compared to other singularity-based models in Stokes flows. Then, in order to model ciliary flows in confined spaces, we extend the point torque solution to a bounded domain between two plane parallel no-slip walls. The flow field is resolved using the method of images and Fourier transforms, and we analyze the role of confinement by comparing the resultant fluid velocity to that of a rotlet near a single wall. Our results suggest that the flow field of a single cilium is not changed significantly by the confinement, even when the distance between the walls is commensurate with the cilium's length.
\end{abstract}

\maketitle

%---------------------------------------------------------------------------
\section{Introduction}

Cilia are microscopic, hair-like appendages found in a diverse array of eukaryotic organisms~\cite{wei2019zero,wei2021measurements}. They are ubiquitous in nature, and perform a variety of functions, including fluid pumping across cell surfaces~\cite{ding2014mixing}, sensing of environmental signals in micro-swimmers~\cite{bloodgood2010sensory}, clearing respiratory airways in mammals~\cite{smith2008modelling}, and capturing food and microbes in aquatic species~\cite{sharp2012multi}. Underlining this functional versatility is the ability of cilia to perform asymmetrical motions, which comprise cyclic power-recovery strokes that interact with the surrounding fluid medium~\cite{wei2019zero,gilpin2020multiscale}. The characteristic length scales of the resulting flows are small, with typical Reynolds numbers of order $10^{-6}-10^{-4}$ \cite{schwartz1997analysis,golestanian2011hydrodynamic,wei2019zero}, so they are well described by the Stokes equations \cite{kim1991microhydrodynamics}. Unsurprisingly, significant advances in the understanding of ciliary hydrodynamics have been made using fundamental solutions of the Stokes equations, including individual Stokeslets \cite{blake1972model,fulford1986muco,mathijssen2016hydrodynamics} or higher order multipole expansions \cite{Vilfan:2012}. Line distributions of Stokeslets have been used, for example, to mimic the interaction of a single cilium with the surrounding fluid~\cite{blake1972model,fulford1986muco,brennen1977fluid}.

While the near-field time-dependent flow field caused by the deformation of individual cilia and flagella has been the subject of intense theoretical and experimental investigation, there are many situations in which the longer-range global flow is of primary interest, particularly in the context of dense arrays and carpets of cilia and their collective motion, e.g., \cite{Bruot2016, Uchida:2010ly}. For large arrays of cilia, present either on motile organisms such as {\it Volvox} \cite{pedley2016squirmers} or {\it Paramecium} \cite{Machemer:1972ys}, or on stationary tissues such as the respiratory tract or on corals \cite{shapiro2014vortical}, the envelope model \cite{Lighthill:1952,blake1971spherical} has proved to be very useful for calculating the motion of individual swimmers \cite{Drescher:2010kx} as well as pairwise interactions and collective dynamics \cite{BrumleyPedley:2019}. This approximates the ciliary array as a no-slip boundary undergoing small-amplitude deformations and derives an effective slip velocity boundary condition on a stationary substrate. The collective dynamics of ciliary beating has also been studied using a minimal model~\cite{vilfan2006hydrodynamic, Niedermayer:2008fk, Friedrich:2012fk, brumley2012hydrodynamic, brumley2014flagellar, brumley2015metachronal}, in which the ciliary tip is modelled as a sphere driven along a circular trajectory [Fig.~\ref{figure1}(a)], so that the flows due to individual cilia are resolved. With the inclusion of either a phase-dependent driving force \cite{uchida2011generic} or compliance in the trajectory shape due to hydrodynamic disturbances \cite{Niedermayer:2008fk}, this model (or `colloidal rotor') provides a good understanding of the emergence of pairwise synchronization as well as metachronal waves \cite{brumley2015metachronal,Brumley2015}, and is in good agreement with experiments \cite{brumley2012hydrodynamic, brumley2015metachronal}. However, both the envelope model and the colloidal rotor model have significant drawbacks when it comes to modelling time-averaged ciliary flows.
 
The envelope model, for example, fails to resolve the vortical flows generated around individual cilia, which play a crucial role in nutrient mixing and exchange~\cite{juan2020multi}. On the other hand, employing the colloidal rotor for finding time-averaged flows of ciliary carpets can consume significant computational time, because the model first solves for an unsteady flow field generated by an orbiting sphere, i.e., the flow field containing a time-evolving singularity, which must then be averaged over the period of oscillations. 

In the present study, we instead use \textit{steady singularities} to capture the time-averaged flow field generated by a cilium. This can easily be scaled up to ciliary carpets using superposition, enabling the computationally-efficient study of flow fields and the corresponding transport of dissolved or particulate matter in different environments. In order to model this flow field, we take a complementary approach to those of \citet{brumley2014flagellar} and \citet{juan2020multi}, but instead capture the time-averaged flow using a point torque. The resulting flow field is thus described by a rotlet, and we compare it to the fields approximated using multiple Stokeslets and the colloidal rotor model. The latter has previously been shown to accurately predict the motion of passive tracers in the fluid around a beating flagellum \cite{brumley2014flagellar, pedley2016squirmers}. Thus, we use the flow field predicted by the colloidal rotor model as the benchmark with which we compare our steady singularity solutions in this paper. We demonstrate the advantages of our approach with respect to accuracy and computational efficiency, compared to previous models and studies, and show that the near-field flow is captured accurately with just one singularity solution. 

Ciliary flows often occur in confined spaces, for example surrounding epithelial cells in the fallopian tube \cite{nutu2007membrane}, near actuated artificial cilia in microfluidic devices \cite{khaderi2011magnetically,den2013microfluidic} and within the mucociliary layer in lung airways \cite{smith2008modelling,juan2020multi}. The latter has been studied recently using Stokeslets~\cite{juan2020multi}, demonstrating the need for adapting flow fields to the presence of parallel boundaries. Changes to the Stokeslet solution by confinement are well documented and typically rely on extensive use of image singularities that can become quite complicated. For example, using Fax\'en's technique \cite{zwanzig1964hydrodynamic,brenner1966stokes}, \citet{de1973low} modified the Stokeslet solution by placing the point force in the middle of two parallel plates. \citet{liron1976stokes} extended this to the case of a point force placed at an arbitrary position in a channel, also using the method of images and Fourier transforms. An alternative analytical form was derived more recently by \citet{mathijssen2016hydrodynamics}, using a recursive series for the image system. In a related study, \citet{hackborn_1990} solved for the asymmetric flow caused by a point torque oriented parallel to two rigid walls, while \citet{van2007stokes} explored the two-dimensional flow field caused by a point torque in a finite rectangular cavity, so that the torque orientation was parallel to the planar wall. 
\citet{dauparas2016flagellar} derived the far-field solution of an arbitrarily oriented point torque placed between two rigid walls while investigating the flagellar flows around bacterial swarms. Following \cite{liron1976stokes}, \citet{fortune_2022} derived the complete solution of the flow field due to a point torque placed in a thin film. However, alterations to the complete rotlet solution (i.e., capturing both the near- and far-field features) for arbitrary orientation of the point torque between parallel rigid walls has not been explored yet. Thus, in this paper, we investigate the influence that plane, parallel no-slip boundaries have on the rotlet solution in general, for arbitrary orientation of the point torque.

This paper is organized as follows. In Section~\ref{S2}, we present mathematical models for the time-averaged flow field of a single cilium. By comparison with the minimal model involving a colloidal rotor, and experimental data of ~\citet{brumley2014flagellar} and \citet{pedley2016squirmers}, we demonstrate that the rotlet-based (point torque) model is superior at capturing the resultant flow above a no-slip wall compared to a range of Stokeslet-based models. By applying similar techniques to~\citet{liron1976stokes}, we then solve for the flow field generated by a point torque placed between two parallel walls in Section~\ref{S3}. Using this framework, we quantify the effects of geometric confinement on the flow induced by individual cilia, and estimate the degree to which the classical rotlet solution near a wall is affected by the (additional) upper boundary. Our conclusions are presented in Section~\ref{S4}. 

%---------------------------------------------------------------------------
\section{Mathematical models of ciliary flow near a wall}\label{S2}

\begin{figure}[htp]
	\begin{center}
		{\includegraphics*[width=\textwidth]{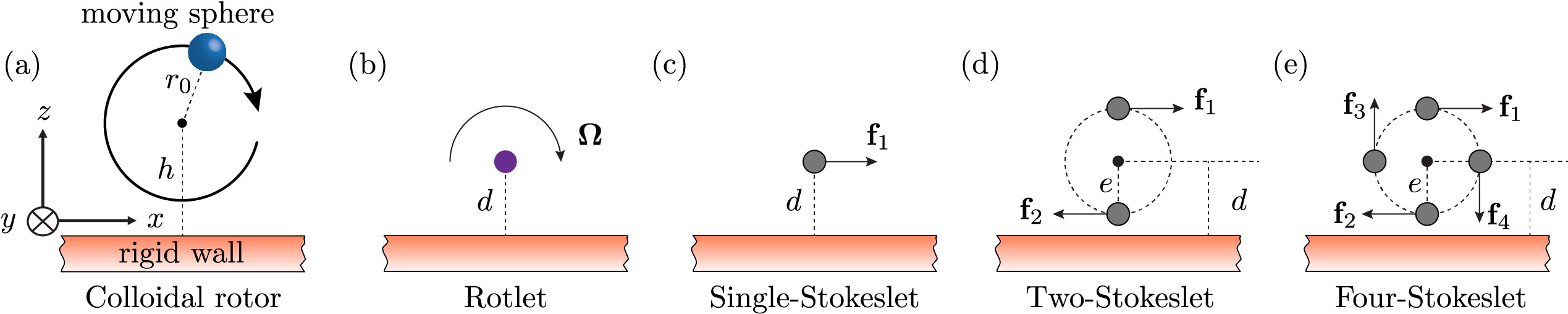}}
	\end{center}\vspace{-0.5cm}
	\caption{Schematic illustration of models used for capturing time-averaged flow generated by an individual cilium. (a) Colloidal rotor mimicking the motion of the ciliary or flagellar tip. (b) The rotlet model; (c) Single-Stokeslet model; (d) Two-Stokeslet model and (e) Four-Stokeslet model. The black arrows in the Stokeslet- and rotlet-based models denote the direction of $q$\textsuperscript{th} point force $\bm{f}_q$ and point torque $\bm{\Omega}$ applied on the fluid domain, respectively, while $d$ and $e$ govern the placement of the singularities with respect to the boundary.}\label{figure1}
\end{figure}

We start by presenting different (singularity-based) models implemented in this paper for capturing the time-averaged flow generated by a cilium above a single, rigid no-slip wall. The fluid flows are described by the Stokes equations \cite{chwang1974hydromechanics} 
\begin{subequations}\label{BigModel}
\begin{eqnarray}
\mu \nabla^2 \bm{u}+\mathcal{F} &= \nabla p,\label{BigModel9a}\\
 \nabla\cdot \bm{u} &= 0,\label{BigModel9b}
\end{eqnarray}\label{ee9}
\end{subequations}
subject to the following boundary conditions on the wall, given by $-\infty<x, y<\infty$ and $z=0$, 
\begin{equation}
\bm{u}=\bm{0}~~\mbox{at}~~z=0,\label{bcBigModel}
\end{equation}
where $\bm{u}=(u_1, u_2, u_3)$ and $p$ denote the fluid velocity and pressure, respectively, and $\mathcal{F}$ corresponds to the forcing that mimics ciliary behavior. Although cilia execute complex changes in their waveform throughout their beating cycle, the instantaneous flow field is well captured by a Stokeslet. To mimic the asymmetry of the power and recovery strokes within a cycle of ciliary beating, we consider several simplified models for the time-averaged flow. We examine the flow field above a no-slip wall generated by a point torque, or a rotlet model (Fig.~\ref{figure1}(b)), as well as a discrete set of one, two, or four point forces, or Stokeslet-based models (Figs.~\ref{figure1}(c-e) respectively) positioned along, and tangential to, a circular trajectory. Optimal values of model parameters for the rotlet and Stokeslet-based models are obtained by comparing the resultant flows to the one found using a colloidal rotor. The latter is the main quantitative benchmark used throughout this paper because it is in agreement with the experimental data of~\citet{pedley2016squirmers} as noted above. The comparison with the colloidal rotor shows that the rotlet provides the best approximation for the time-averaged flow generated by a cilium using the steady singularities approach. 

\subsection{The colloidal rotor model}

The colloidal rotor model assumes that the ciliary tip is a sphere of radius $a$ translating in a closed circular loop of radius $r_0$ and driven by a constant angular velocity of magnitude $\omega$ pointing into the page in Fig.~\ref{figure1}(a). Hence, the model predicts a time-periodic ciliary flow. In order to compute the time-averaged component of the velocity vector for the individual cilia at a position $\bm{x}=(x,y,z)$, we apply the time averaging as
\begin{eqnarray}
u_i^s=\frac{\omega}{2\pi}\int_{0}^{2\pi/\omega} \frac{1}{8\pi\mu} S_{ij}(\bm{X}(t),\bm{x}) \text{F}_j(t)~dt.
\label{e2}
\end{eqnarray}
Here, $\bm{X}(t)=(x_s(t),y_s(t),z_s(t))=(r_0\sin{\omega t},0,h+r_0\cos{\omega t})$ is the trajectory of the sphere, and the force exerted on the fluid of viscosity $\mu$ due to the moving sphere is $F_j(t)=\gamma_{jk}X_k'(t)=6\pi\mu a\left(\delta_{jk}+\frac{9a}{16 z_s(t)}\left(\delta_{jk}+\delta_{j3}\right)\right)X_k'(t)$, where $\delta_{jk}$ is the Kronecker delta. Thus, $\mathcal{F}=\textbf{F}(t) \delta (\bm{x}-\bm{X}(t))$ in Eq.~\eqref{BigModel9a}. Furthermore, $S_{ij}$ is the second-order Green's tensor of the Stokes flow due to a point force above the wall (i.e., a Stokeslet) \cite{blake1971note}, given by 
\begin{eqnarray}\label{Stokeslet}
S_{ij}(\bm{X}(t),\bm{x})=\frac{\delta_{ij}}{r}+\frac{r_ir_j}{r^3}-\bigg(\frac{\delta_{ij}}{R}+\frac{R_iR_j}{R^3}\bigg)+2z_s(t) \rho_{jk3}\biggl[\frac{\partial}{\partial R_k}\bigg(\frac{z_s(t) R_i}{R^3}-\bigg[\frac{\delta_{i3}}{R}+\frac{R_iR_3}{R^3}\bigg]\bigg)\bigg],
\end{eqnarray}
where $\rho_{jk3}=\delta_{j\alpha}\delta_{\alpha k}-\delta_{j3}\delta_{3 k}$ for $\alpha \in \{1,2\}$, 
$\bm{x}=(x, y, z)$, $\bm{r}=(r_1,r_2,r_3)=(x-x_s(t),y-y_s(t),z-z_s(t))$, $\bm{R}=(R_1,R_2,R_3)=(x-x_s(t),y-y_s(t),z+z_s(t))$ and indices $\{i,j,k\} \in \{1,2,3\}$.

\subsection{Stokeslet-based models}\label{S22}

Next, we consider three models for capturing the time-averaged ciliary flow based on the increasing number of point forces in a Stokes flow. This is designed to mimic the force applied by the cilia at an increasing number of points throughout the beating cycle.

\textbf{\textit{Single-Stokeslet model.}} We model the time-averaged ciliary flow as generated by a point force $\bm{f}_1=|\bm{f}|\bm{\hat{e}}_x$, of magnitude $|\bm{f}|$ and orientation $\bm{\hat{e}}_x$ (where $\bm{\hat{e}}_{\varphi}$ henceforth corresponds to the unit vector along the $\varphi$-direction) located at $\bm{x}_{s_1}=(x_{s_1},y_{s_1},z_{s_1})=(0,0,d)$ above the wall (see Fig.~\ref{figure1}(c)), i.e. $\mathcal{F} =\bm{f}_1\delta(\bm{x}-\bm{x}_{s_1})$ in Eq.~\eqref{BigModel9a}. This acts as the power stroke pushing the fluid forward along the $+x$ direction and is designed to capture the part of the beating cycle with the strongest flow.

\textbf{\textit{Two-Stokeslet model.}} In this model, we assume that the time-averaged ciliary flow corresponds to the flow generated by two point forces $\bm{f}_1=|\bm{f}|\bm{\hat{e}}_x$ and $\bm{f}_2=-|\bm{f}|\bm{\hat{e}}_x$ of equal magnitude $|\bm{f}|$, but acting in opposite directions parallel to the wall. The forces are positioned at $\bm{x}_{s_1}=(x_{s_1},y_{s_1},z_{s_1})=(0,0,d+e)$ and $\bm{x}_{s_2}=(x_{s_2},y_{s_2},z_{s_2})=(0,0,d-e)$, respectively, so that they are $2e$ apart and the midpoint between them is at a distance $d$ above the wall, see Fig.~\ref{figure1}(d). The model, therefore, accounts for both the power stroke (which pushes the fluid forward) and the recovery stroke (which pushes the fluid backward), resulting in $\mathcal{F} =\bm{f}_1\delta(\bm{x}-\bm{x}_{s_1})+\bm{f}_2\delta(\bm{x}-\bm{x}_{s_2})$ in Eq.~\eqref{BigModel9a}. Although these two forces are considered to have the same magnitude, the resulting fluid disturbance will not be the same, owing to hydrodynamic screening due to the wall.

\textbf{\textit{Four-Stokeslet model.}} Finally, the time-averaged ciliary flow is modelled by placing four point-forces $\bm{f}_1=|\bm{f}|\bm{\hat{e}}_x$, $\bm{f}_2=-|\bm{f}|\bm{\hat{e}}_x$, $\bm{f}_3=|\bm{f}|\bm{\hat{e}}_z$ and $\bm{f}_4=-|\bm{f}|\bm{\hat{e}}_z$ of equal magnitude $|\bm{f}|$ at $\bm{x}_{s_1}=(x_{s_1},y_{s_1},z_{s_1})=(0,0,d+e)$, $\bm{x}_{s_2}=(x_{s_2},y_{s_2},z_{s_2})=(0,0,d-e)$, $\bm{x}_{s_3}=(x_{s_3},y_{s_3},z_{s_3})=(-e,0,d)$ and $\bm{x}_{s_4}=(x_{s_4},y_{s_4},z_{s_4})=(e,0,d)$, respectively, as shown in Fig.~\ref{figure1}(e), so that $\mathcal{F} =\sum_{i=1}^4\bm{f}_i\delta(\bm{x}-\bm{x}_{s_i})$ in Eq.~\eqref{BigModel9a}. The distance between the horizontally and vertically aligned point forces is $2e$, and the geometrical centre of the force assembly is a distance $d$ above the wall. The ciliary power strokes are captured by the point forces oriented along the $+x$ and $+z$ directions (i.e., $\bm{f}_1$ and $\bm{f}_3$), whereas the recovery strokes are captured by the point forces oriented along the $-x$ and $-z$ directions (i.e., $\bm{f}_2$ and $\bm{f}_4$). 

The general expression for the fluid velocity component at $\bm{x}=(x,y,z)$ obtained using Stokeslet models above is:
\begin{align}
&u_i=\frac{dx_i}{dt}=\frac{1}{8\pi\mu} \sum^{N}_{q=1} S_{ij}(\bm{x}_{s_q},\bm{x}) f_{qj},\label{e4}
\end{align}
where $N \in \mathbb{N}$ denotes the number of point forces used in the model (i.e. $N=1,2~\text{or}~4$), $\bm{x}_{s_q}=(x_{s_q},y_{s_q},z_{s_q})$ is the position of the $q$\textsuperscript{th} point force $\bm{f}_q$ of magnitude $|\bm{f}|$ and $S_{ij}(\bm{x}_{s_q},\bm{x})$ is the Green's tensor, which corresponds to the $q$\textsuperscript{th} point force above the wall and is similar to Eq.~\eqref{Stokeslet}, but does not contain time-evolving variables, 
\begin{eqnarray}\label{Stokeslet2}
&&S_{ij}(\bm{x}_{s_q},\bm{x})=\frac{\delta_{ij}}{r}+\frac{r_ir_j}{r^3}-\bigg(\frac{\delta_{ij}}{R}+\frac{R_iR_j}{R^3}\bigg)+2z_{s_q} \rho_{jk3}\biggl[\frac{\partial}{\partial R_k}\bigg(\frac{z_{s_q} R_i}{R^3}-\bigg[\frac{\delta_{i3}}{R}+\frac{R_iR_3}{R^3}\bigg]\bigg)\bigg].
\end{eqnarray}
Here the vector $\bm{r}=(r_1,r_2,r_3)=(x-x_{s_q},y-y_{s_q},z-z_{s_q})$, the vector $\bm{R}=(R_1,R_2,R_3)=(x-x_{s_q},y-y_{s_q},z+z_{s_q})$, and the remaining notation is the same as in Eqs.~\eqref{e2}-\eqref{Stokeslet}. Note that for $N>1$, the velocity of the fluid (\ref{e4}) is obtained by the superposition of velocities that correspond to all Stokeslets in the domain.

\subsection{The rotlet model} \label{S21}

Finally, we model the time-averaged ciliary flow as generated by a point torque $\bm{\Omega}=|\bm{\Omega}|~\hat{\bm{e}}_y$ of magnitude $|\bm{\Omega}|$ and orientation $\hat{\bm{e}}_y$ positioned at $\bm{x}_r=(0,0,d)$, as shown in Fig.~\ref{figure1}(b), so that $\mathcal{F} = 4\pi\mu \nabla \times (\mathbf{\Omega}\mathbf{\delta}(\bm{x}-\bm{x}_r))$  \cite{chwang1974hydromechanics}. This generates a flow with velocity components given by 
\begin{align}
&u_i=\frac{dx_i}{dt}=\frac{1}{8\pi\mu} A_{ij}(\bm{x}_{r},\bm{x}) \Omega_j,\label{e5}
\end{align}
where $A_{ij}$ is the Green's tensor due to a point torque (i.e., rotlet) above the wall in Stokes flow \cite{blake1974fundamental}, given by 
\begin{eqnarray}
A_{ij}(\bm{x}_r,\bm{x})&=&\frac{\epsilon_{ijk} r_k}{r^3}-\frac{\epsilon_{ijk}R_k}{R^3}+2d\epsilon_{kj3} \bigg(\frac{\delta_{ik}}{R^3}-\frac{3R_iR_k}{R^5}\bigg)+\frac{6\epsilon_{kj3}R_iR_kR_3}{R^5}. \label{GreenRot}
\end{eqnarray}
Here $\bm{x}=(x,y,z)$, $r=|\bm{r}|$, $\bm{r}=(r_1,r_2,r_3)=(x,y,z-d)$, $\bm{R}=(R_1,R_2,R_3)=(x,y,z+d)$, $\epsilon_{ijk}$ is the Levi-Civita symbol; as before, $\mu$ is the fluid viscosity and the indices $i,j,k \in \{1,2,3\}$. 

\subsection{Model optimizations}\label{S23}
The flow generated by a distribution of steady singularities as detailed above is optimized to best capture the time-averaged flow field obtained using the colloidal rotor, i.e. Eq.~\eqref{e2}, which accurately mimics the behavior of a eukaryotic flagellum isolated from the colonial alga \textit{Volvox carteri} \cite{pedley2016squirmers}. These flagella, found in water with $\mu=10^{-3}\, \text{Pa}\, \text{s}$, are of average length $\langle l_c \rangle = 19.9~\mu$m and beat with a period $T\sim 1/33$~s, suggesting that the reference scales for length, velocity, force, and torque in Eqs.~\eqref{e4}-\eqref{GreenRot} are $l_c$, $U_t=2\pi l_c/T$, $f_r=\mu l_c^2/T$ and $\Omega_{r}=f_{r} l_c$, respectively. In the case of the colloidal rotor, the experimental data for the \textit{Volvox} flagellum is captured well for $a/l_c=0.25$, $h/l_c=0.5$ and $r_0/l_c=0.25$~\cite{pedley2016squirmers}. We, therefore, optimize the steady singularity-based models by adjusting $d$, $e$, $|\bm{f}|$ and $|\bm{\Omega}|$ and comparing the resultant velocity field to that obtained from Eq.~\eqref{e2} using the above parameter values for the colloidal rotor model. The numerical optimization is carried out outside a semicircular region $\Delta$ of radius $2l_c$, chosen to mask the flow singularities. However, we have found that the outcome of our optimization was qualitatively insensitive to this choice. Thus, the fluid velocity remains regular within the optimization domain, while important near- and far-field features of the ciliary flow are nonetheless still captured. For the remainder of this section, we present the results in the plane $y=0$.

We start by computing 
\begin{equation}
 \text{RD} =\bigg|\frac{U^s-U^f}{U^s}\bigg|,\label{e6}
\end{equation}
which we refer to as the relative difference, and then the global mean of this quantity 
\begin{equation}
\langle\text{RD}\rangle =\frac{1}{m}\sum_m\text{RD} = \frac{1}{m}\sum_m\bigg|\frac{U^s-U^f}{U^s}\bigg|,\label{e6a}
\end{equation}
where $m$ is the number of specified uniformly distributed grid points in the region of interest $x \cup z \in [-L/2l_c,\; L/2l_c]\cup[0,\; H/l_c] =[-5,\;5]\cup[0,\;10]$ and $y=0$, but outside the region $\Delta$; we used a very fine mesh, so that $m$ was up to $10^{6}$ across all model comparisons (see also Fig.~\ref{figure3}(b)). Here, $U^s=|\bm{u}^s|=\sqrt{\sum_{i=1}^{3} (u^s_i)^2}$ and $U^f=|\bm{u}|=\sqrt{\sum_{i=1}^{3} u^{2}_i}$ are the magnitudes of time-averaged velocity field obtained using the colloidal rotor (i.e., Eq.~\eqref{e2}) and the particular steady singularity-based model (i.e., Eq.~\eqref{e4} and \eqref{e5}), respectively. The expression in Eq.~\eqref{e6a} is minimized by varying parameters $d\in I_d$ and $|\mathbf{\Omega}| \in I_{\Omega}$ for the rotlet model, and $d\in I_d$, $e \in I_e$ and $|\textit{\textbf{f}}| \in I_{f}$ for the Stokeslet-based models (in fact, in the case of the single Stokeslet model, $\langle\text{RD}\rangle$ is only a function of $d$ and $|\textit{\textbf{f}}|$). Thus, the rotlet and the single Stokeslet models are the most straightforward to optimize. The intervals $I_d$, $I_{\Omega}$, $I_{f}$ and $I_{e}$ are selected using the reference scales above, so that $I_d/l_c,\;I_e/l_c,\;I_f/(\mu l^2_c/T)\;\text{and}\;I_{\Omega}/(\mu l^3_c/T) \sim \mathcal{O}(1)$, and $d$, $e$ $|\textit{\textbf{f}}|$, $|\mathbf{\Omega}|$ are varied by taking $n_d$, $n_e$, $n_f$, $n_{\Omega}$ steps on these intervals, respectively. That is, the geometric parameters associated with the singularity-based models are considered to be commensurate with those of the colloidal oscillator model \cite{pedley2016squirmers}.

\begin{figure}[!t]
	\begin{center}
		{\includegraphics[width=\textwidth]{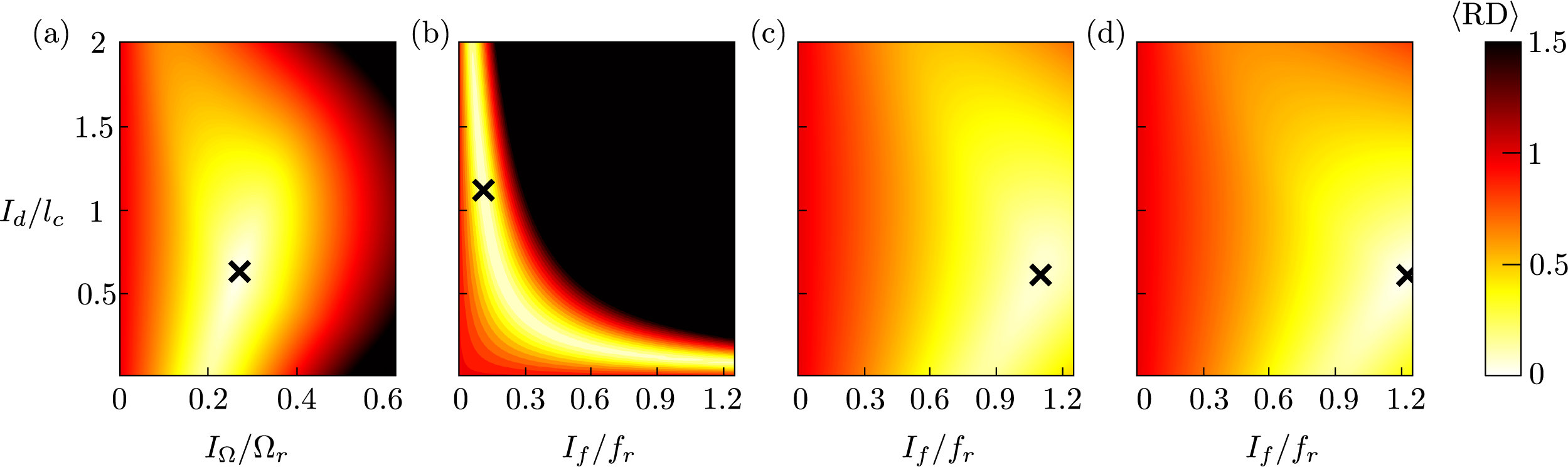}}
	\end{center}
	\caption{The heat map of the mean relative difference $\langle \text{RD} \rangle$ (a) in $(I_d/l_c,I_{\Omega}/\Omega_r)$ plane for the rotlet model, (b) in $(I_d/l_c, I_f/f_r)$ plane for the single-Stokeslet model, (c) in $(I_d/l_c, I_f/f_r)$ plane at $e/l_c=0.059$ for the two-Stokeslet model and (d) in $(I_d/l_c, I_f/f_r)$ plane at $e/l_c=0.054$ for the four-Stokeslet model. The minimum value in the heat map is indicated using the black cross. The map is generated using $n_d=100$, $n_f=100$, $n_{\Omega}=100$ and $n_e=50$ points on the intervals $I_d/l_c$, $I_f/f_r$, $I_{\Omega}/\Omega_r$ shown in (a)-(d), and $I_e/l_c=[0,0.25]$, respectively.} \label{figure2}
\end{figure} 

The results of this optimization procedure are illustrated in Fig.~\ref{figure2}, where we show the computed $\langle\text{RD}\rangle$ using a heat map for different values of parameters on the optimization grid. For example, the mean relative difference, Eq.~\eqref{e6a}, for the rotlet model, shown in Fig.~\ref{figure2}(a), is plotted on the $(I_d/l_c, I_{\Omega}/\Omega_r)$ plane. The minimum value of $\langle\text{RD}\rangle$ is attained at the point marked with a black cross in Fig.~\ref{figure2}(a) (and in the remaining panels of Fig.~\ref{figure2}), and the corresponding parameter values are cited in Table~\ref{table1}. The optimization parameter space for the single Stokeslet model is also two-dimensional; the corresponding heat map of $\langle\text{RD}\rangle$ in $(I_d/l_c, I_{f}/f_r)$ plane is illustrated in Fig.~\ref{figure2}(b). However, the optimization parameter space for the two and four-Stokeslet models is three-dimensional, and so in Figs.~\ref{figure2}(c) and~\ref{figure2}(d) we plot a cross-section of that space in the $(I_d/l_c, I_{f}/f_r)$ plane at a fixed value of $e=0.059l_c$ and $e=0.054l_c$, respectively, for which $\langle \text{RD} \rangle$ attains the minimum (optimal) value, see Table~\ref{table1}. Note that for these two models, we obtain the same optimal values of $d$ due to the symmetry of the Stokeslet spatial distribution. 

\begin{table}[htp!]
    \centering
    \begin{tabular}{|l|l|}
    \hline
        Model type & Optimal parameters values \\ \hline
        rotlet model & $d/l_c=0.629$, $|\mathbf{\Omega}|/\Omega_r=0.271$\\ \hline
        single-Stokeslet model & $d/l_c=1.111$, $|\textit{\textbf{f}}|/f_r=0.113$ \\
        \hline
        two-Stokeslet model & $d/l_c= 0.609$, $e/l_c=0.059$, $|\textit{\textbf{f}}|/f_r=1.098$\\
        \hline
        four-Stokeslet model & $d/l_c= 0.609$, $e/l_c=0.054$, $|\textit{\textbf{f}}|/f_r=1.224$
        \\ \hline
    \end{tabular}
\caption{Results of the optimization procedure for different models.}\label{table1}
\end{table}

\begin{figure}[h!]
	\begin{center}
	{\includegraphics*[width=\textwidth]{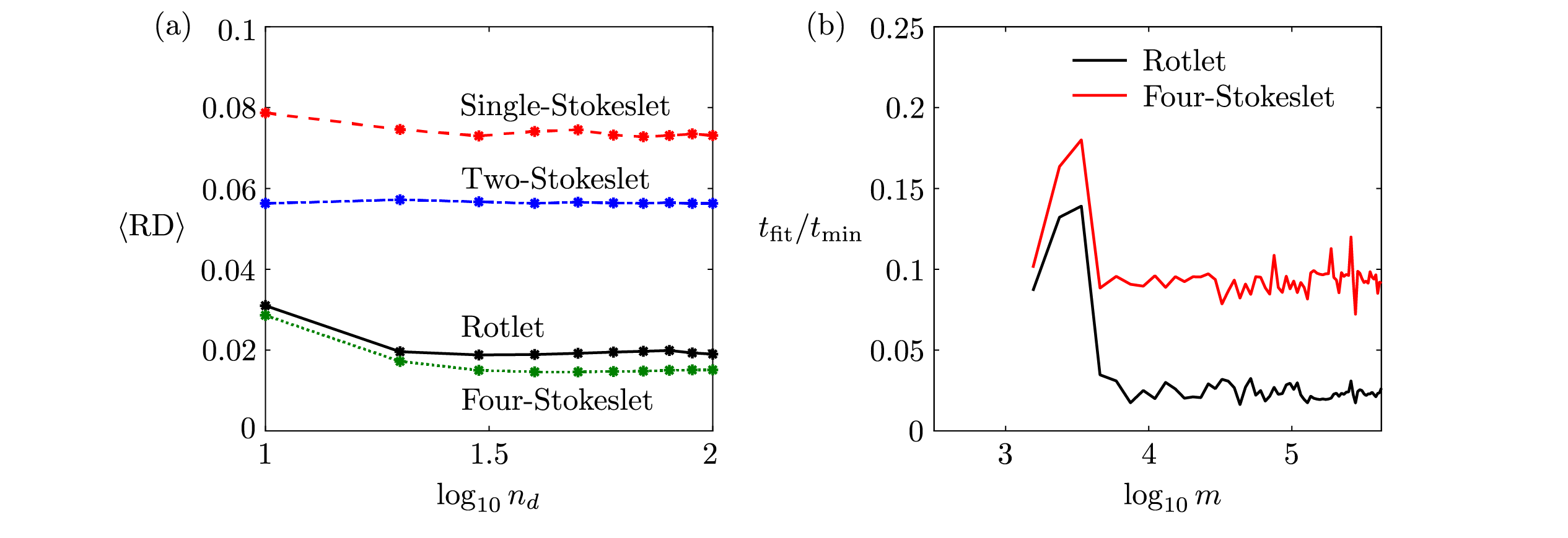}}
	\end{center}
	\caption{(a) The mean relative difference $\langle \text{RD} \rangle$ as a function of $\mbox{log}_{10} n_d$ for different steady singularity models. (b) The ratio of computational times taken to compute the time-averaged flow field using the steady singularity models $(t_{\text{fit}})$ and the colloidal rotor model indicated with the legend $(t_{\text{min}})$ as a function of $\mbox{log}_{10} (m)$, where $m$ is the number of computational grid points.}\label{figure3}
\end{figure}

The results presented in Table~\ref{table1} are quite insensitive to the number of points chosen for the optimization process, e.g., $n_d$ on $I_d$. To illustrate this, in Fig.~\ref{figure3}(a) we show how the minimal value of $\langle \text{RD} \rangle$ evolves with the logarithm of $n_d$. For all steady singularity-based models, the minimum of $\langle \text{RD} \rangle$ is approximately the same for $n_d\gtrsim50$, and the location of this minimum in the optimization parameter space is approximately the same. Also notable in Fig.~\ref{figure3}(a) is that Stokeslet-based models with fewer point forces result in a larger value of $\langle \text{RD} \rangle$ compared to the rotlet model, suggesting that the latter outperforms them. However, adding additional point forces in the Stokeslet-based models reduces the value of $\langle \text{RD} \rangle$. 

We leave further accuracy comparison between these models to \S\ref{ModelComparison} and focus instead on the computational efficiency of obtaining the time-averaged flow. In Fig.~\ref{figure3}(b), the relative computational time for the steady singularity and the colloidal rotor models is plotted as a logarithm of the number of grid points in the computational domain. The plot suggests that the time-averaged flow field is between 7 and 50 times faster to compute using the rotlet model, depending on the number of grid points than the colloidal rotor model. Similarly, using the four-Stokeslet model is between 5 and 10 times faster than the colloidal rotor model. This is unsurprising, given that the colloidal rotor computations involve evaluating and then averaging a time-periodic velocity field (see Eq.~\eqref{e2}). Even more significant is the fact that the rotlet model requires fewer parameters to fit the time-averaged flow as compared to the four-Stokeslet model, and is also up to 5 times quicker according to Fig.~\ref{figure3}(b).

\subsection{Model comparisons}\label{ModelComparison}

\begin{figure}[h!]
	\begin{center}
		{\includegraphics*[width=\textwidth]{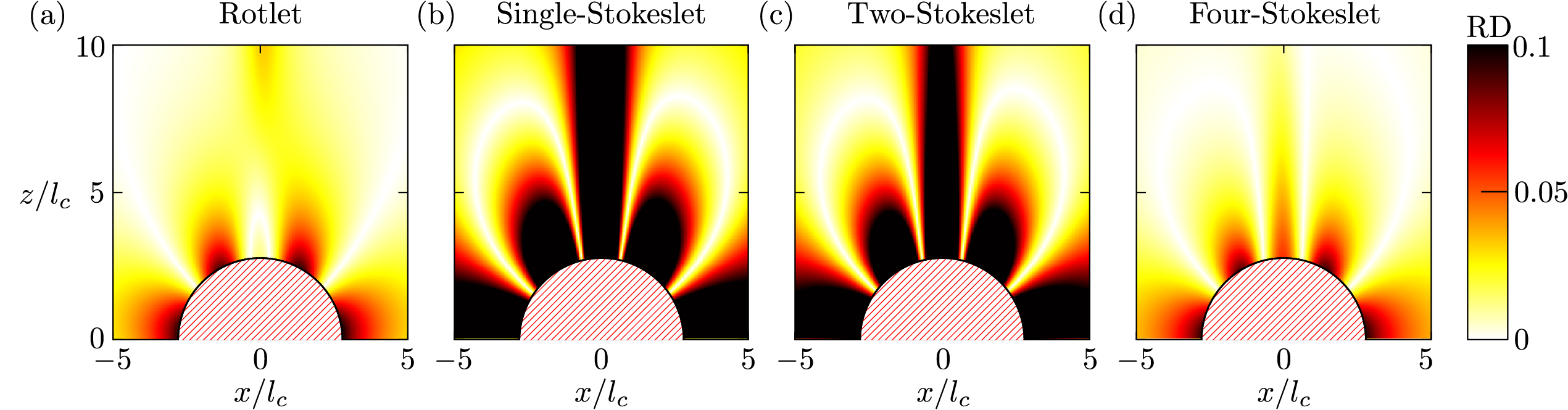}}
	\end{center}\vspace{-0.5cm}
	\caption{The relative difference ($\mbox{RD}$) obtained using (a) the rotlet model, (b) the single-Stokeslet model, (c) the two-Stokeslet model, and (d) the four-Stokeslet model in the window of length $L/l_c=10$ and height $H/l_c=10$ at optimal parameters values listed in Table~\ref{table1}. The red shaded semicircle in (a)-(d) corresponds to the region $\Delta$ excluded from the calculations of RD.} \label{figure4}
\end{figure}

\begin{figure}[htp!]
	\begin{center}
		{\includegraphics*[width=\textwidth]{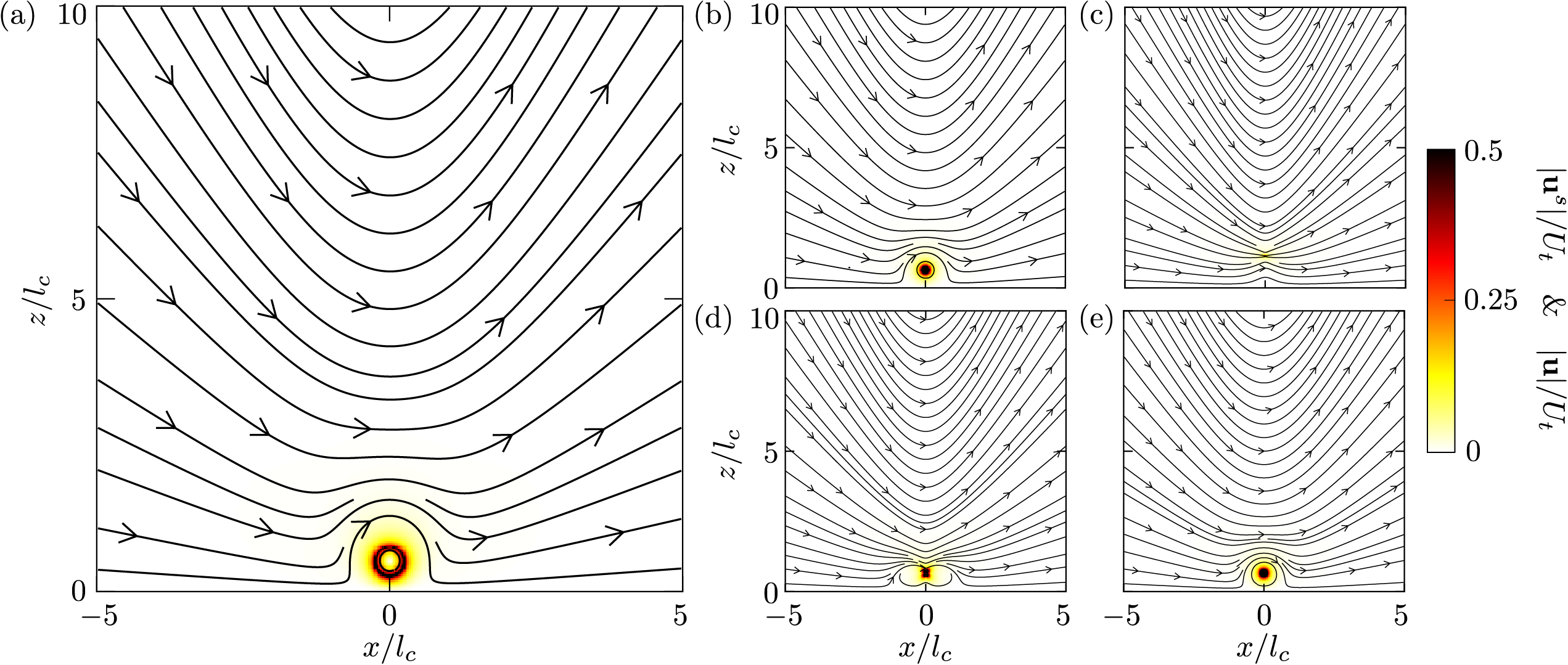}}
	\end{center}\vspace{-0.5cm}
	\caption{Streamlines and velocity magnitude associated with the (a) time-averaged colloidal rotor, (b) the rotlet model, (c) the single-Stokeslet model, (d) the two-Stokeslet model, and (e) the four-Stokeslet model, all shown in the window of length $L/l_c=10$ and height $H/l_c=10$ at optimal parameters values listed in Table~\ref{table1}. Streamlines are shown with lines that have arrows to indicate the flow direction.}\label{figure5}
\end{figure}

We continue the comparison between the different steady singularity-based models by plotting the heat map of the relative difference RD (i.e. Eq.~\eqref{e6}) in the plane $y=0$ for the optimal values of the parameters listed in Table~\ref{table1}. These spatial plots of RD are shown in Fig.~\ref{figure4}. As evident from Fig.~\ref{figure4}, the Stokeslet-based models with a larger number of point forces outperform (on accuracy) the single Stokeslet model. 
The heat map of the relative difference for the rotlet model looks similar to the four-Stokeslet model, and we find that the predictions for the time-averaged flow field obtained with both models are much closer to those of the colloidal rotor. This observation is further confirmed in Fig.~\ref{figure5}, where the flow from the colloidal rotor (panel (a)) is shown alongside the optimized flow fields for various singularity models (panels (b)-(e)). While the differences between the models are less pronounced in the far field ($z/l_c \gtrsim 7$) -- where the time-averaged velocity decays as $1/r^2$ in all models -- there are significant differences between the models in the near-field ($z/l_c \lesssim 5$), with the rotlet and four-Stokeslet models closely capturing the flow of the colloidal oscillator.

\begin{figure}[h!]
	\begin{center}
	{\includegraphics[width=\textwidth]{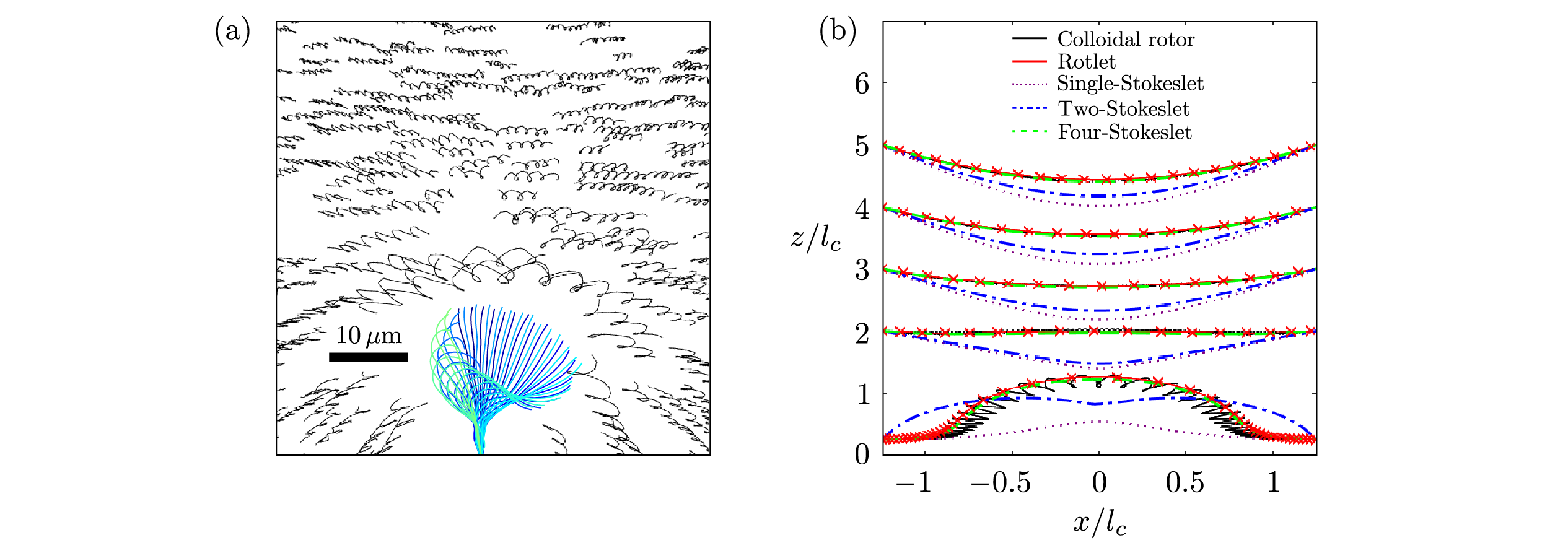}}
	\end{center}
	\vspace{-0.5cm}
	\caption{(a) From \cite{pedley2016squirmers}: a sequence of instantaneous shapes of an isolated eukaryotic flagellum held by a micropipette (blue waveform); the resulting trajectories of passive microspheres are shown in black. (b) Trajectories of passive tracers, initially positioned at $x_0/l_c=-1.26$ and $z_0/l_c=0.25, 2, 3, 4, 5$, respectively, obtained using the different singularity-based models in a window of length $L/l_c=2.5$ and height $H/l_c=7$. As discussed in Section~\ref{S23}, the model parameters are chosen for comparison with \cite{pedley2016squirmers}.}\label{figure6}
\end{figure}

Next, in Fig.~\ref{figure6} we also compare the flow field generated by the singularity-based models with the experimental tracks of microspheres in the vicinity of a single eukaryotic flagellum \cite{pedley2016squirmers}. As illustrated in Fig.~\ref{figure6}(a), the microspheres follow undulating trajectories due to the asymmetric beating of the flagellum, with undulations increasing in amplitude closer to the flagellum.
Similar trajectories are computed using the colloidal rotor for various initial positions $(x_0,z_0)$, see Fig.~\ref{figure6}(b). If the steady singularities-based models are employed instead, the trajectory is non-oscillatory, see also Fig.~\ref{figure6}(b). It is evident that the time-averaged trajectories in both the near- and far-field predicted using the Stokeslet-based models improve when additional point forces are used, the closest agreement being for the four-Stokeslet case. However, the rotlet model accurately captures the time-averaged trajectories of passive tracers with the smallest number of fitting parameters, and is in close agreement with published experimental trajectories \cite{pedley2016squirmers}. Taken together, these results demonstrate that the simple representation of flagellar dynamics using a point torque accurately captures the time-averaged flow in both the near- and far-field.

%---------------------------------------------------------------------------
\newpage

\section{Modelling time-averaged ciliary flow between two parallel rigid walls}\label{S3}

In this section, we consider the hydrodynamic effects of confining a point torque between two plane parallel rigid walls. The goals of this are to determine the fluid flows generated by beating cilia in confined geometries -- for example, in biological tissues and microfluidic devices -- but also to determine the conditions under which confinement may be appropriately neglected.

\subsection{Flow due to a point torque between two parallel rigid walls}\label{S31}

\begin{figure}[htp!]
	\begin{center}
          {\includegraphics*[width=\textwidth]{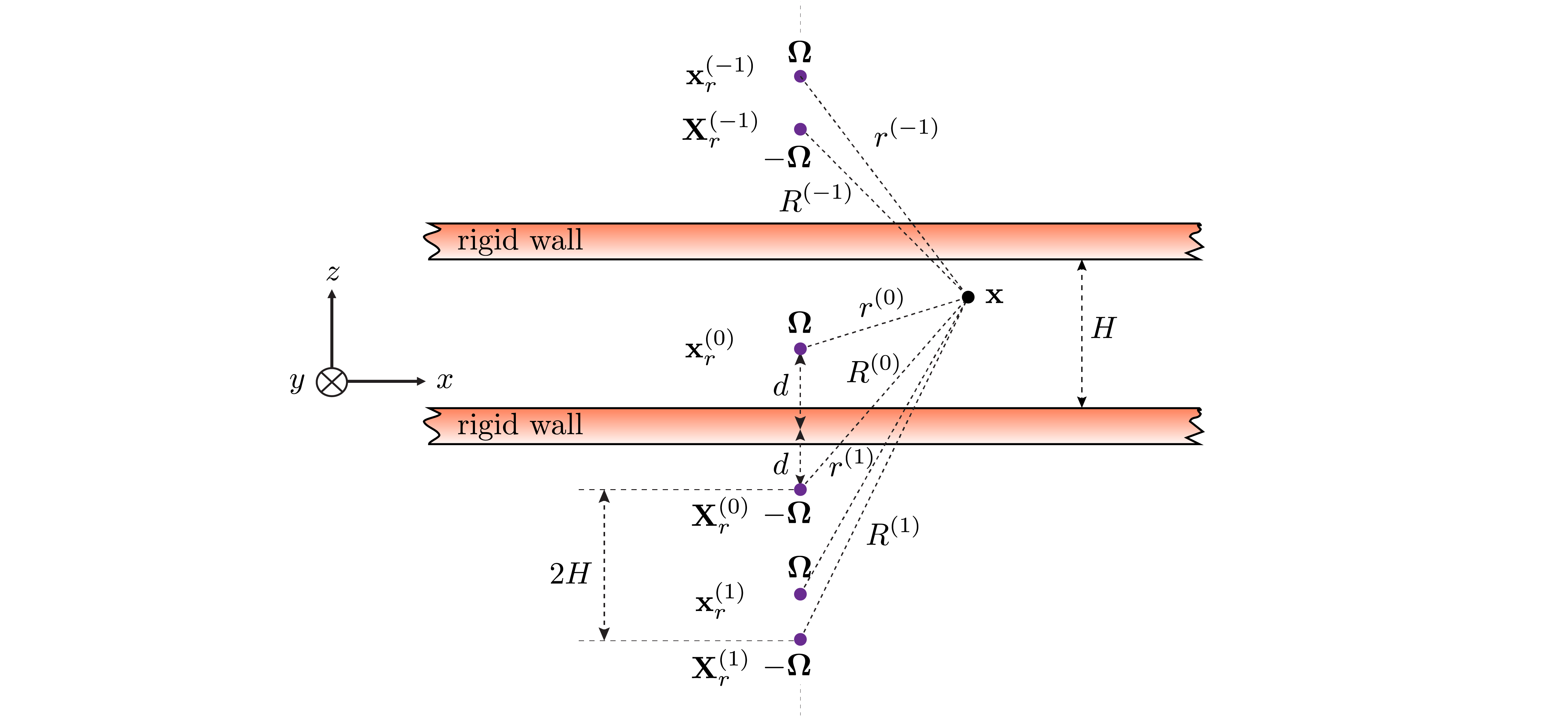}}
	\end{center}\vspace{-0.5cm}
	\caption{Schematic diagram showing the point torque, $\mathbf{\Omega}$, situated between the two parallel walls at $z=0$ and $z=H$, and its image system. The flow field is derived for arbitrary orientation of the point torque.} \label{figure7}
\end{figure}

The flow field generated by a point torque and given by Eq.~\eqref{e5} must be modified when situated in a bounded flow domain. To derive a model for the time-averaged ciliary flow between two parallel walls separated by a distance $H$ (i.e., lying within the region $-\infty<x, y<\infty$ and $z=0, H$, see Fig.~\ref{figure7}), we solve Eqs.~\eqref{BigModel}-\eqref{bcBigModel} for $\mathcal{F} = 4\pi\mu \nabla \times (\mathbf{\Omega}\mathbf{\delta}(\bm{x}-\bm{x}_r^{(0)}))$, where $\mathbf{\Omega}=\Omega_1\hat{\mathbf{e}}_x + \Omega_2\hat{\mathbf{e}}_y + \Omega_3\hat{\mathbf{e}}_z$ is a point torque positioned at $\bm{x}_r =\bm{x}_r^{(0)}=(0,0,d)$, but with the additional no-slip boundary condition on the top wall located at $z=H$. That is,
\begin{equation}
\bm{u}=\bm{0}~~\mbox{at}~~z=H.\label{bcBigModel2}
\end{equation}
We follow a parallel approach to that of \citet{liron1976stokes}, who solved the equivalent problem, but for the flow generated by a point force. Unlike \citet{hackborn_1990}, this method solves for the flow field generated by a point torque of arbitrary orientation between the rigid walls. We commence by introducing images of the point torque, located with respect to the parallel walls as shown in Fig.~\ref{figure7}, which, once the principle of superposition is applied, ensures that the no-penetration boundary conditions are satisfied. 

The images are built as follows: we reflect the solution for the flow induced by a point torque of strength $|\mathbf{\Omega}|$ in an unbounded domain with respect to the lower wall by placing the image at $\bm{X}_r^{(0)}=(0, 0, -d)$ as was done when deriving the model in \S\ref{S21}, and with the respect to the upper wall by placing an additional image at $\bm{X}_r^{(-1)}=(0, 0, 2H-d)$. However, these images then induce a non-zero normal velocity component $u_3$ on the opposite walls, so further reflections of the source at $\bm{x}_r^{(0)}$ and its images located at $\bm{X}_r^{(0)}$ and $\bm{X}_r^{(-1)}$ are needed with respect to both walls. Following the same procedure as in the first step, we built a sequence of reflections by positioning point torques, all of equal magnitudes, at the following positions:
\begin{eqnarray}
\bm{x}_r^{(n)}=(0, 0,d-2nH)~~~\mbox{and}~~~\bm{X}_r^{(n)}=(0, 0,-d-2nH),~~~\mbox{for}~~~n \in \mathbb{Z}. \label{ee12}
\end{eqnarray}
Here the superscripts $n\geq 0$ pertain to images below the lower wall, and superscripts $n<0$ correspond to the images above the upper wall. The resulting flow field is given by 
\begin{eqnarray}
 u_i=S_{ij}\Omega_j,\label{ee11}
\end{eqnarray}
where 
\begin{equation}
 S_{ij}=\frac{\varepsilon_{ijk}}{8\pi\mu} \sum^{\infty}_{n=-\infty} \bigg\{\frac{r_k^{(n)}}{(r^{(n)})^3}-\frac{R_k^{(n)}}{(R^{(n)})^3}\bigg\}
 \label{ee13}
\end{equation}
is the singular Green's kernel obtained by superposing the rotlet solution in an unbounded domain located at $\bm{x}_r^{(0)}$ and all of its images. Here, $r^{(n)}=|\bm{r}^{(n)}|$ and $R^{(n)}=|\bm{R}^{(n)}|$ denote the amplitudes of vectors $\bm{r}^{(n)}=\bm{x}-\bm{x}_r^{(n)}=(x,y,z-d+2nH)$ and $\bm{R}^{(n)}=\bm{x}-\bm{X}_r^{(n)}=(x,y,z+d+2nH)$, where $\bm{x}=(x,y,z)$ is a general point in the flow domain between the two parallel walls, and, as before, indices $i,j,k \in \{1,2,3\}$ correspond to the $x$, $y$, $z$ directions respectively.
Using the definition of the Lipschitz integral \cite{liron1976stokes,watson1922treatise},
\begin{equation}
\int_0^{\infty} J_0(\tau \lambda)~e^{-|\theta|\lambda}~d\lambda = \frac{1}{\sqrt{\tau^2+\theta^2}},    \label{e14}
\end{equation}
where $\theta \in \{z-d+2nH, z+d+2nH\}$ and $\tau=\sqrt{x^2+y^2}$, we obtain the following expressions for the geometric sum 
\begin{eqnarray}
   \sum^{\infty}_{n=-\infty} \bigg\{\frac{1}{r^{(n)}}-\frac{1}{R^{(n)}}\bigg\}=\begin{cases}
    \displaystyle 2 \int_0^{\infty} J_0(\tau \lambda) \frac{\sinh{\lambda (H-d)}}{\sinh{\lambda H}}~\sinh{\lambda z}~d\lambda~~\mbox{for}~~z<d,\vspace{0.3cm}\\
    \displaystyle 2 \int_0^{\infty} J_0(\tau \lambda) \frac{\sinh{\lambda d}}{\sinh{\lambda H}}~\sinh{\lambda (H-z)}~d\lambda~~\mbox{for}~~z\geq d.
    \end{cases}\label{ee14a}
\end{eqnarray}
Here and elsewhere, $J_{\nu}$ corresponds to the Bessel function of the first kind, where its order $\nu \in \{0,1\}$.
The integral approximation for the infinite series in Eq.~\eqref{ee13} can thus be found by taking the derivative of Eq.~\eqref{ee14a} with respect to $r_k$:
\begin{eqnarray}
    \sum^{\infty}_{n=-\infty} \bigg\{\frac{r_k^{(n)}}{(r^{(n)})^3}-\frac{R_k^{(n)}}{(R^{(n)})^3}\bigg\}=\begin{cases}
    \displaystyle 2 \delta_{k \alpha}\frac{r_{\alpha}}{\tau} \int_0^{\infty}\lambda J_1(\tau \lambda) \frac{\sinh{\lambda (H-d)}}{\sinh{\lambda H}}~\sinh{\lambda z}~d\lambda \vspace{0.3cm}\\
    \displaystyle \qquad \qquad - 2 \delta_{k3}\int_0^{\infty}\lambda J_0(\tau \lambda) \frac{\sinh{\lambda (H-d)}}{\sinh{\lambda H}}~\cosh{\lambda z}~d\lambda
    ~\quad~\mbox{for}~\quad~z<d,\vspace{0.5cm}\\
    \displaystyle 2 \delta_{k \alpha}\frac{r_{\alpha}}{\tau} \int_0^{\infty}\lambda J_1(\tau \lambda) \frac{\sinh{\lambda d}}{\sinh{\lambda H}}~\sinh{\lambda (H-z)}~d\lambda \vspace{0.3cm}\\
    \displaystyle \qquad \qquad + 2 \delta_{k3} \int_0^{\infty}\lambda J_0(\tau \lambda) \frac{\sinh{\lambda d}}{\sinh{\lambda H}}~\cosh{\lambda (H-z)}~d\lambda
    ~\quad~\mbox{for}~\quad~z\geq d,
    \end{cases}\label{ee14b}
\end{eqnarray}
so that the kernel $S_{ij}$ in Eq.~\eqref{ee11} can be rewritten as
\begin{subequations}
\begin{eqnarray}
S_{\alpha \beta}&=&\begin{cases}\displaystyle-\frac{\varepsilon_{\alpha \beta 3}}{4\pi \mu} \int_0^{\infty}\lambda J_0(\tau \lambda) \frac{\sinh{\lambda (H-d)}}{\sinh{\lambda H}}~\cosh{\lambda z}~d\lambda~~\mbox{for}~~z<d,
\vspace{0.3cm}\\
\displaystyle \frac{\varepsilon_{\alpha \beta 3}}{4\pi \mu}\int_0^{\infty} \lambda J_0(\tau \lambda) \frac{\sinh{\lambda d}}{\sinh{\lambda H}}~\cosh{\lambda (H-z)} d\lambda~~\mbox{for}~~z\geq d,
\end{cases}\label{e15a}\\
S_{3 \alpha}&=&\begin{cases}\displaystyle \frac{\varepsilon_{3 \alpha \beta}}{4\pi \mu} \frac{r_{\beta}}{\tau} \int_0^{\infty}\lambda J_1(\tau \lambda) \frac{\sinh{\lambda (H-d)}}{\sinh{\lambda H}}~\sinh{\lambda z}~d\lambda~~\mbox{for}~~z<d,
\vspace{0.3cm}\\
\displaystyle \frac{\varepsilon_{3 \alpha \beta}}{4\pi \mu} \frac{r_{\beta}}{\tau} \int_0^{\infty} \lambda J_1(\tau \lambda) \frac{\sinh{\lambda d}}{\sinh{\lambda H}}~\sinh{\lambda (H-z)} d\lambda~~\mbox{for}~~z\geq d,
\end{cases}\label{e15b}\\ \vspace{0.5cm}
S_{i3}&=&\begin{cases}\displaystyle \frac{\varepsilon_{i 3\alpha}}{4\pi \mu} \frac{r_{\alpha}}{\tau} \int_0^{\infty}\lambda J_1(\tau \lambda) \frac{\sinh{\lambda (H-d)}}{\sinh{\lambda H}}~\sinh{\lambda z}~d\lambda~~\mbox{for}~~z<d,
\vspace{0.3cm}\\
\displaystyle \frac{\varepsilon_{i 3\alpha}}{4\pi \mu} \frac{r_{\alpha}}{\tau} \int_0^{\infty} \lambda J_1(\tau \lambda) \frac{\sinh{\lambda d}}{\sinh{\lambda H}}~\sinh{\lambda (H-z)} d\lambda~~\mbox{for}~~z\geq d.
\end{cases}\label{e15c}
\end{eqnarray}\label{e15}
\end{subequations}
As before, the indices $\alpha$, $\beta \in \{1,2\}$. This implies that along the walls at $z=0$, $H$, the singular Green's kernel $S_{ij}$, given by Eq.~\eqref{e15}, is equal to
\begin{subequations}\label{SingKerBoun}
\begin{eqnarray}
S_{\alpha\beta}&=&\begin{cases}
\displaystyle -\frac{\varepsilon_{\alpha\beta 3}}{4\pi\mu}\int_0^{\infty} \lambda J_0(\tau \lambda) \frac{\sinh{\lambda (H-d)}}{\sinh{\lambda H}}~d\lambda~~~\mbox{for}~~~z= 0,
\vspace{0.1cm}\\
\displaystyle  \frac{\varepsilon_{\alpha\beta 3}}{4\pi\mu} \int_0^{\infty} \lambda J_0(\tau \lambda) \frac{\sinh{\lambda d}}{\sinh{\lambda H}}~d\lambda~~~\mbox{for}~~~z= H,
\end{cases}\\
S_{3 \alpha}&=&0~~\mbox{and}~~S_{i3}=0~~\mbox{at}~~z=0,~H,
\end{eqnarray}\label{e16}
\end{subequations}
containing only two non-zero components, $S_{12}$ and $S_{21}$. Therefore, by construction, the flow field due to an arbitrary orientated point torque given by Eq.~\eqref{ee11} in combination with Eq.~\eqref{e15}, satisfies the Stokes equations (\ref{BigModel}), and impermeability boundary conditions on both walls, i.e., $u_3=0$ at $z=0$ and $z=H$, but in general results in non-zero tangential velocity components $u_1$ and $u_2$ along the boundaries for the point torque oriented parallel to these walls. However, when the point torque is oriented perpendicular to the walls, the no-slip and impermeability conditions are satisfied. In this case, the flow field (given by Eq.~\eqref{e15c} or equivalent series Eq.~\eqref{ee13}) resembles the classical rotlet solution in an unbounded domain \cite{chwang1974hydromechanics}, and decays as $1/r^{2}$ both in the near- and far-field everywhere except near the walls, where it is exactly zero due to the superposition. In order to build a solution that satisfies the no-slip boundary conditions for arbitrary orientation of a point torque, we once again use the principle of superposition, and, as in Ref~\cite{liron1976stokes}, introduce auxiliary kernels $F_{ij}$ and $P_j$, such that the flow field due to the rotlet $\mathbf{\Omega}$ between two parallel walls is given by 
\begin{eqnarray}
    u_i=(S_{ij}+F_{ij})\Omega_j\quad\text{and}\quad p=P_j\Omega_j.\label{ee17}
\end{eqnarray}
Substituting the above into the Eq.~\eqref{ee9} results in:
\begin{subequations}
\begin{eqnarray}
   (\mu \nabla^2 F_{ij}-\nabla_i P_j) \Omega_j &=& 0,\label{ne1a} \\   
   \nabla_iF_{ij}\Omega_j &=& 0,\label{ne1b}
\end{eqnarray}\label{ne1}
\end{subequations}
because the singular solution $S_{ij}\Omega_{ij}$ satisfies Eq.~\eqref{ee9}. When $\bm{\Omega}\neq \bm{0}$, the system of equations~\eqref{ne1} has a non-trivial solution if:
\begin{subequations}\label{newStokes}
\begin{eqnarray}
   \mu \nabla^2 F_{ij} &=&\nabla_i P_j ,\label{ee18a} \\   
   \nabla_iF_{ij}&=& 0.\label{ee18b}
\end{eqnarray}\label{ee18}
\end{subequations}
These equations are subject to the boundary conditions 
\begin{equation}
    F_{ij}=-S_{ij}~~~\mbox{at}~~~z=0, H, \label{ee19}
\end{equation}
which ensures that the flow field in Eq.~\eqref{ee17} satisfies both the no-penetration and no-slip boundary conditions on both walls, i.e., Eqs.~\eqref{bcBigModel} and \eqref{bcBigModel2}. This also implies that the pressure kernel $P_j$, which corresponds to the auxiliary solution, satisfies Laplace's equation $\nabla^2 P_j =0.$

We solve for the auxiliary kernels $F_{ij}$ and $P_j$ by applying a two-dimensional Fourier transform of the form
\begin{equation}
\hat{\mathcal{L}}_{ij}=\int_{-\infty}^{\infty}\int_{-\infty}^{\infty} \mathcal{L}_{ij}e^{\mathrm{i}(\zeta_1x+\zeta_2y)}dxdy, 
\end{equation}
to Eq.~\eqref{newStokes} ~\cite{liron1976stokes,mathijssen2016hydrodynamics},
which results in
\begin{subequations}
\begin{eqnarray}
-\mathrm{i}\zeta_{\alpha} \delta_{\alpha i} \hat{P}_j + \delta_{i3} \frac{\partial \hat{P}_j}{\partial z} &=&\mu \bigg(\frac{\partial^2}{\partial z^2}-\zeta^2\bigg)\hat{F}_{ij},\label{ee21a}\\
-\mathrm{i}\zeta_{\alpha} \hat{F}_{\alpha j}+\frac{\partial \hat{F}_{3j}}{\partial z}&=&0,\label{ee21b}
\end{eqnarray}\label{ee21}
\end{subequations}
where $\zeta=|\bm{\zeta}|=\sqrt{\zeta_1^2+\zeta_2^2}$ is the lateral distance of the point $\bm{\zeta}=(\zeta_1,\zeta_2)$ from the point torque in Fourier space. Similarly, the associated boundary conditions are obtained by taking the Fourier transform of Eq.~\eqref{ee19} combined with Eq.~\eqref{SingKerBoun}, resulting in
\begin{subequations}
\begin{eqnarray}
   4\pi\mu \hat{F}_{\alpha\beta} &=& \displaystyle \varepsilon_{\alpha\beta3} \frac{\sinh{\zeta (H-d)}}{\sinh{\zeta H}}~~~\mbox{at}~~~z=0, \label{ee22a}\\
    4\pi\mu \hat{F}_{\alpha\beta} &=& \displaystyle -\varepsilon_{\alpha\beta3} \frac{\sinh{\zeta d}}{\sinh{\zeta H}}~~~\mbox{at}~~~z=H, \label{ee22b}\\
    4\pi\mu \hat{F}_{3\alpha} &=&0~~~\mbox{and}~~~4\pi\mu \hat{F}_{3\alpha}=0~~~\mbox{at}~~~z=0,H.\label{ee22c}
\end{eqnarray}\label{ee22}
\end{subequations}
Solving the transformed Laplace's equation for the pressure kernel, we obtain  
\begin{equation}
    \hat{P}_j=D_j \sinh{\zeta (H-z)}+E_j \cosh{\zeta (H-z)},\label{ee24}
\end{equation}
where $D_j$ and $E_j$ are constant coefficients. Substituting this into Eq.~\eqref{ee21a} allows us to determine the transformed auxiliary kernel $\hat{F}_{ij}$:
\begin{eqnarray}
   &\mu \hat{F}_{ij}=B_{ij} \sinh{\zeta(H-z)}+ C_{ij} \cosh{\zeta(H-z)}+\bigg\{D_j\delta_{i3}+E_j \delta_{\alpha i} \bigg(\displaystyle\frac{\mathrm{i}\zeta_{\alpha}}{\zeta}\bigg)\bigg\}z \sinh{\zeta(H-z)} \nonumber\\
    &+\bigg\{E_j\delta_{i3}+D_j \delta_{\alpha i} \bigg(\displaystyle \frac{\mathrm{i}\zeta_{\alpha}}{\zeta}\bigg) \bigg\} (z-H) \cosh{\zeta (H-z)}, \label{ee25}
\end{eqnarray}
where $B_{ij}$ are $C_{ij}$ are also constant coefficients. The relationships between the different coefficients are obtained from the continuity equation, by substituting Eq.~\eqref{ee25} into Eq.~\eqref{ee21b}, collecting the coefficients of $\sinh{\zeta (H-z)}$ and $\cosh{\zeta (H-z)}$ and equating the two sets to zero:
\begin{subequations}
\begin{eqnarray}
    E_j &=& \zeta H D_j+\zeta B_{3j}+\mathrm{i}\zeta_{\beta} C_{\beta j},\label{ee26a}\\
    D_j &=& -\zeta H E_j+\zeta C_{3j}+\mathrm{i}\zeta_{\beta} B_{\beta j}. \label{ee26b}
\end{eqnarray} \label{ee26}
\end{subequations}
Finally, by combining the above with boundary conditions in Eq.~\eqref{ee22}, we are able to determine
all coefficients: 
\begin{subequations}
\begin{eqnarray}
    B_{ij}&=&  \frac{\displaystyle  \varepsilon_{i j 3}}{4\pi} \frac{\cosh{\zeta d}}{\sinh{\zeta H}}+\delta_{\alpha i}\bigg(\frac{\mathrm{i}\zeta_{i} H}{\zeta} \frac{\cosh{\zeta H}}{\sinh{\zeta H}}\bigg) D_j+ \delta_{i3} \frac{H \cosh{\zeta H}}{\sinh{\zeta H}} E_j,\label{ee27a}\\
    C_{ij}&=&-\frac{\varepsilon_{ij3}}{4\pi} \frac{\sinh{\zeta d}}{\sinh{\zeta H}},\label{ee27b}\\
    D_i&=&\frac{\mathrm{i}\zeta_{j}~\varepsilon_{ij3}}{4\pi}~\frac{\zeta H \cosh{\zeta(H-d)}-\cosh{\zeta d} \sinh{\zeta H}}{\sinh^2{\zeta H}-(\zeta H)^2},\label{ee27c}\\
    E_i&=&\frac{\mathrm{i}\zeta_{j}~\varepsilon_{ij3}}{4\pi}~\frac{\sinh{\zeta d}\sinh{\zeta H}-\zeta H \sinh{\zeta(H-d)}}{\sinh^2{\zeta H}-(\zeta H)^2}.\label{ee27d}
\end{eqnarray} \label{ee27}
\end{subequations}
Implementing Eq.~\eqref{ee27} fixes the transformed auxiliary solutions $\hat{P}_j$ and $\hat{F}_{ij}$, and, once the inverse Fourier transforms are taken, enables us to write the components of the auxiliary Green's kernel for velocity and pressure as:
\begin{subequations}
\begin{eqnarray}
    F_{\alpha \beta}&=&\frac{\varepsilon_{k \beta 3}}{4\pi\mu} \bigg[\delta_{\alpha k} \int_0^{\infty} J_0(\tau \zeta)~\phi(\zeta)~d\zeta -\bigg[\frac{\delta_{\alpha k}}{\tau}
    -\frac{r_{\alpha} r_k}{\tau^3}\bigg]\int_0^{\infty} J_1(\tau \zeta)~\chi(\zeta)~d\zeta - \frac{r_{\alpha} r_k}{\tau^2}\int_0^{\infty} J_1'(\tau \zeta)~\chi(\zeta)~d\zeta \bigg], \label{ee28a}\\
    F_{3\alpha}&=& \frac{\varepsilon_{k\alpha 3}}{4\pi\mu}~\frac{r_k}{\tau}\int^{\infty}_0 J_1(\tau \zeta)~\kappa(\zeta)~d\zeta,\label{ee28b} \\
    F_{i3}&=&0, \label{ee28c}\\
     P_j&=&\frac{\varepsilon_{kj3}}{4\pi}~\frac{r_k}{\tau} \int_0^{\infty} J_1(\tau \zeta)~\psi(\zeta)~d\zeta,\label{ee28d}
\end{eqnarray}\label{ee28}
\end{subequations}
where
\begin{eqnarray}
    \phi(\zeta)&=&\displaystyle \frac{\zeta \sinh{\zeta(H-z-d)}}{\sinh{\zeta H}}, \nonumber\\
    \chi(\zeta)&=&\displaystyle \frac{-\zeta}{\sinh^2{\zeta H}-(\zeta H)^2} \bigg[ \zeta H z \cosh{\zeta(d-z)}-z \sinh{\zeta H} \cosh{\zeta (H-z-d)} \nonumber\\
        &-&   \frac{\zeta H^2 \cosh{\zeta(H-d)}\sinh{\zeta z}}{\sinh{\zeta H}}+H\cosh{\zeta d}\sinh{\zeta z}\bigg],\nonumber\\
     \kappa(\zeta)&=&\displaystyle \frac{-\zeta^2}{\sinh^2{\zeta H}-(\zeta H)^2}\bigg[ \zeta H z \sinh{\zeta(d-z)}-z \sinh{\zeta H} \sinh{\zeta (H-z-d)} \nonumber\\
    \displaystyle  &+&\frac{\zeta H^2 \sinh{\zeta(H-d)}\sinh{\zeta z}}{\sinh{\zeta H}}-H\sinh{\zeta d}\sinh{\zeta z}\bigg],  \nonumber \\
    \psi(\zeta)&=&\frac{-\zeta^2}{\sinh^2{\zeta H}-(\zeta H)^2} \bigg[\zeta H \sinh{\zeta(H-z)}-\sinh{\zeta H} \sinh{\zeta(H-z-d)}\bigg].\nonumber
\end{eqnarray}
 Superposing the above auxiliary kernel in Eq.~\eqref{ee28} with the singular kernel Eq.~\eqref{e15} yields the flow field due to a point torque located between parallel walls via Eq.~\eqref{ee17}. 

\subsection{Effects of confinement}

\begin{figure}[htp!]
	\begin{center}
	    {\includegraphics*[width=\textwidth]{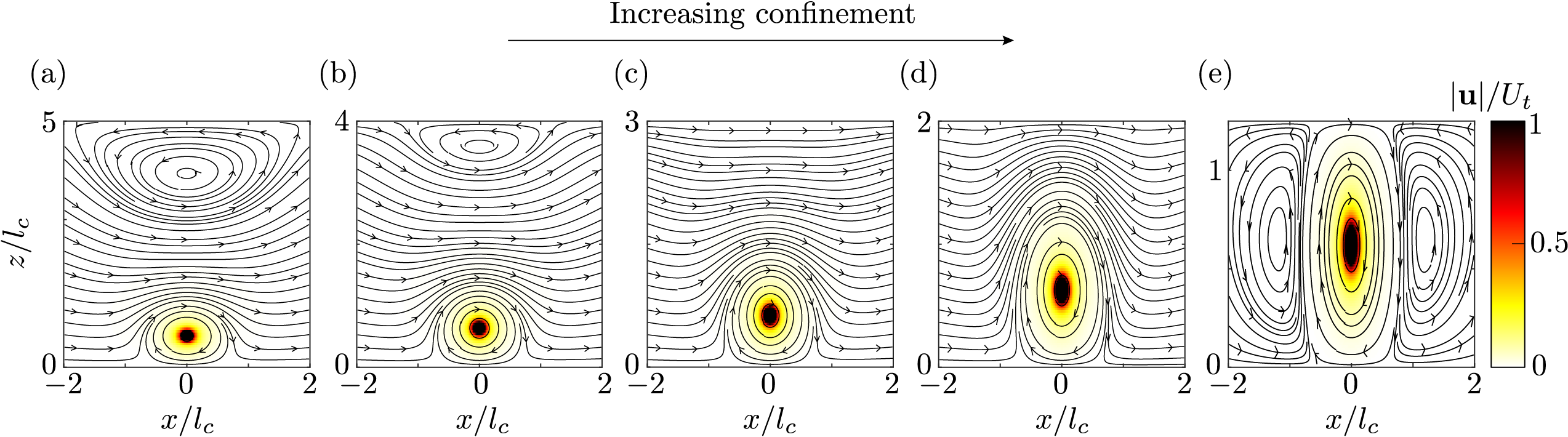}}
	\end{center}
	\caption{Heat map of the fluid velocity magnitude $|\bm{u}|/U_t$ due to a point torque positioned at a distance $d=0.629l_c$ above the bottom wall. The top and bottom walls are separated by a distance of (a) $H=5l_c$, (b) $H=4l_c$, (c) $H=3l_c$, (d) $H=2l_c$ and (e) $H=1.26l_c$. The point torque is oriented parallel to both walls ($\mathbf{\Omega}=|\mathbf{\Omega}|\hat{\mathbf{e}}_y$). The remaining parameters are listed in Table~\ref{table1}. Streamlines are shown with lines that have arrows to indicate the flow direction.}\label{figure8}
\end{figure}

We investigate the effect of confinement separately for the point torque oriented parallel or perpendicular to rigid walls. First, we validate our findings with the analysis of \citet{hackborn_1990} for the point torque oriented parallel to walls ($\mathbf{\Omega}=|\mathbf{\Omega}|\hat{\mathbf{e}}_y$). Figure~\ref{figure8} displays the fluid velocity magnitude and streamlines obtained using Eq.~\eqref{ee17} in the $y=0$ plane, for a range of  distances $H$ between the parallel walls (calculations have been performed as discussed in Appendix~\ref{SecB}). In the vicinity of the point torque, located at $x=y=0$ and $z=d/l_c$, fluid rotates clockwise in accordance with the rotlet orientation. However, for sufficiently large $H$~($H\gtrsim 3.4l_c$, see also Figs.~\ref{figure8}(a) and (b)), a secondary counter-rotating weak vortex forms near the upper wall. As $H$ decreases, the flow along this wall increases gradually~(Figs.~\ref{figure8}(c) and (d)) until $H=2d$ (i.e., twice the distance between point torque and the lower wall), where the point torque generates weak, secondary, counter-rotating vortices in the middle of the channel adjacent to and on either side of the strong, primary vortex (see Fig.~\ref{figure8}(e)). These findings are in line with the flow field observed by \citet{hackborn_1990} for a point torque parallel to the rigid walls.

\begin{figure}[!h]
	\begin{center}
	    {\includegraphics*[width=\textwidth]{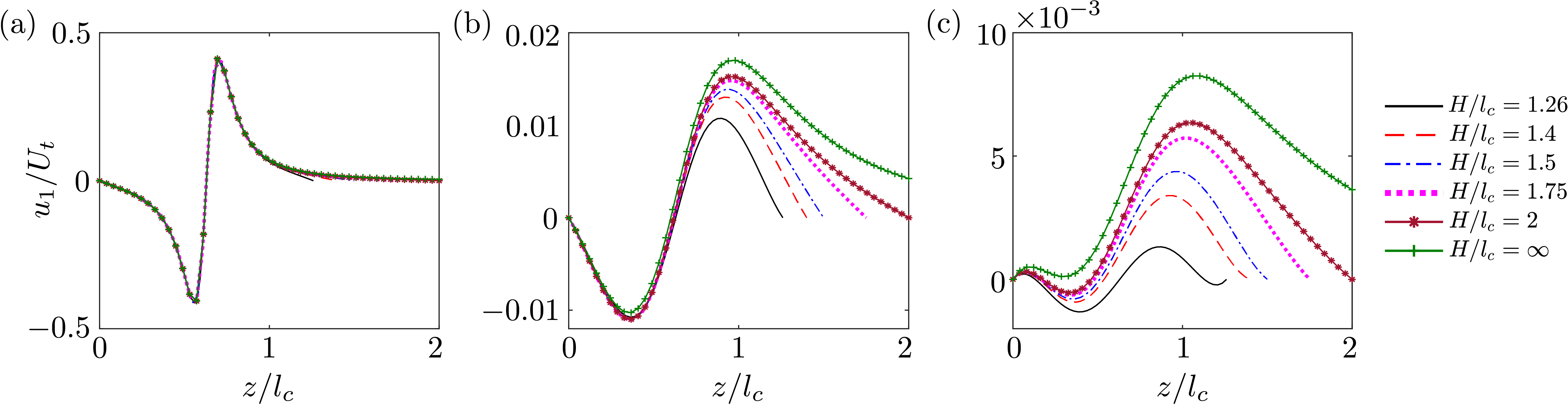}}
	\end{center}
	\caption{Variation of the scaled $x$-component of the velocity vector, i.e. $u_1/U_t$, with scaled $z/l_c$ in the $y=0$ plane for various distances between parallel walls calculated at (a) $x=0.1l_c$, (b) $x=0.5l_c$ and (c) $x=0.75l_c$ from the position of the point torque at $x_r=y_r=0$ and $z_r=d/l_c$ (where $d=0.629 l_c$) oriented parallel to walls ($\mathbf{\Omega}=|\mathbf{\Omega}|\hat{\mathbf{e}}_y$). Model parameters are listed in Table~\ref{table1}.}\label{figure9}
\end{figure}

Following \cite{liron1976stokes}, for the point torque oriented parallel to walls, we also investigate the role of confinement on the profiles of the velocity components. Figure~\ref{figure9} shows the $x$-component of the velocity vector $u_1$ as a function of $z$ at three different $x$ positions in the channel (all with $y=0$), for various values of $H$. There is surprisingly little change to the velocity profiles with varying levels of confinement, with the maximum value of $u_1$ shifting towards larger $z$ and increasing overall for larger $H$. As expected, the $u_1$ profiles in the bounded domains (with finite $H$) approach the profile in the semi-bounded domain as the value of $H \to \infty$. (In Appendix~\ref{A}, we also check that the solution in the semi-bounded domain can be derived from Eq.~\eqref{ee17}).

\begin{figure}[!h]
	\begin{center}
	    {\includegraphics*[width=17.5cm]{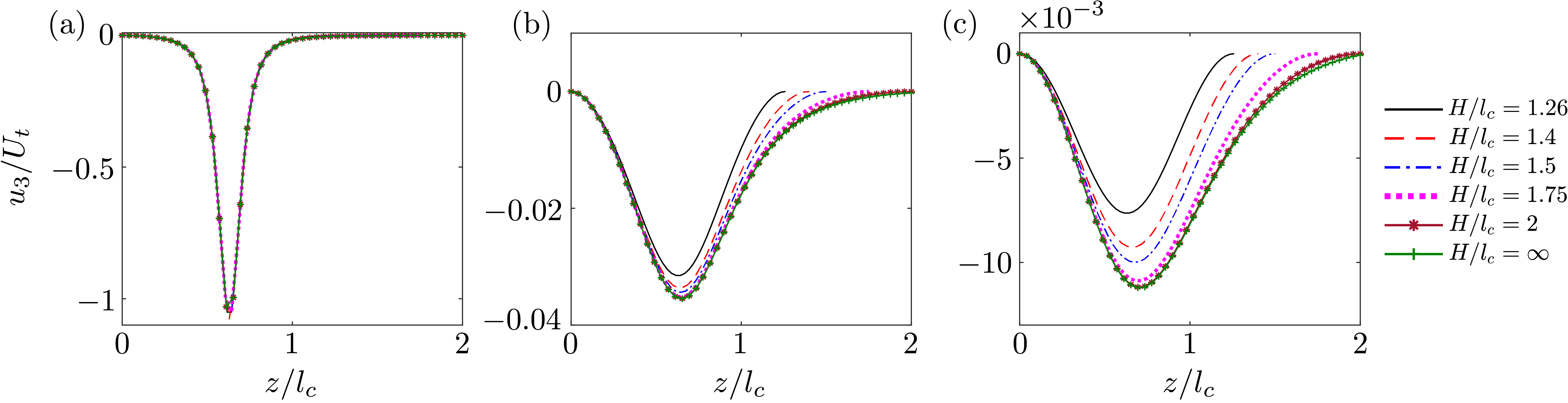}}
	\end{center}
	\caption{The same profiles as in Fig.~\ref{figure9}, but for the $z$-component of the fluid velocity.}\label{figure10}
\end{figure}

While the corresponding $z$-component velocity profiles $u_3$, shown in Fig.~\ref{figure10}, are qualitatively different from the profiles in Fig.~\ref{figure9}, the same conclusions can be made about how they vary with the height of the confinement. The greater the distance, $H$, between the parallel walls, the larger the maximum value of $|u_3|$, and the closer the profiles resemble the predictions in the semi-bounded domain.

\begin{table}[!h]
    \centering
    \begin{tabular}{|c|c|c|c|c|c|}
    \hline
        PD$_x^{\parallel}$ & $H/l_c=1.26$ & $H/l_c=1.4$ & $H/l_c=1.5$ & $H/l_c=1.75$ & $H/l_c=2$\\ \hline
    $x/l_c=0.1$ & $0.13$ & $1.69$ & $0.11$ & $1.55$ & $0.31$\\ \hline
        $x/l_c=0.5$ & $36.84$ & $23.33$ & $18.44$ & $12.73$ & $10.39$\\ \hline
        $x/l_c=0.75$ & $84$ & $58.77$ & $46.98$ & $30.58$ & $23.04$\\
        \hline
    \end{tabular}
\caption{Percentage difference (PD$_x$) between the maximum value of the fluid velocity $x$-component in the semi-bounded and the bounded domains found using Eq.~\eqref{PD_for} for different $x$ and $H$. The corresponding velocity profiles are shown in Fig.~\ref{figure9}. Model parameters are given in Table~\ref{table1}.}\label{table2}
\end{table}

We quantify these observations by studying the percentage difference between the maxima found for each of the two velocity components in the bounded and semi-bounded domains, respectively:
\begin{eqnarray}
   \text{PD}_x^{\parallel}=\bigg|\frac{|u_1(\bm{x}_{\parallel})|_{sb}-|u_1(\bm{\xi}_{\parallel})|_b}{|u_1(\bm{x}_{\parallel})|_{sb}}\bigg| \times 100\%~~~\mbox{and}~~~   \text{PD}_z^{\parallel}=\bigg|\frac{|u_3(\bm{x}_{\parallel})|_{sb}-|u_3(\bm{\xi}_{\parallel})|_{b}}{|u_3(\bm{x}_{\parallel})|_{sb}}\bigg| \times 100\%,\label{PD_for}
\end{eqnarray}
where $\bm{\xi}_{\parallel}$ and $\bm{x}_{\parallel}$ are the positions at which the magnitude of velocity components in the bounded domain (denoted by subscript ``$b$'') and semi-bounded domain (denoted by subscript ``$sb$''), respectively, attain the maximum values at fixed $x$ and $H$. The computed $\text{PD}_x^{\parallel}$ and $\text{PD}_z^{\parallel}$ are shown in Tables~\ref{table2} and ~\ref{table3} for the profiles in Figs.~\ref{figure9} and~\ref{figure10}, respectively. These calculations confirm the earlier observations about the role of the upper wall in changing the flow field of a point torque. The largest percentage difference of 84\% is calculated for $H=1.26l_c$ at $x=0.75l_c$, but the resulting velocity is $10^{-3}$ times $U_t$, which suggests that these effects would be difficult to measure in an experiment. Nevertheless, the cumulative effect of flows due to multiple cilia between parallel walls is expected to be more significant. 

\begin{table}[!h]
    \centering
    \begin{tabular}{|c|c|c|c|c|c|}
    \hline
       PD$_z^{\parallel}$ & $H/l_c=1.26$ & $H/l_c=1.4$ & $H/l_c=1.5$ & $H/l_c=1.75$ & $H/l_c=2$\\ \hline
       $x/l_c=0.1$ & $1.95$ & $4.55$ & $1.93$ & $1.95$ & $0.001$\\ \hline
        $x/l_c=0.5$ & $11.29$ & $5.39$ & $3.12$ & $0.79$ & $0.11$\\ \hline
         $x/l_c=0.75$ & $31.93$ & $17.32$ & $10.89$ & $3.02$ & $0.30$\\
        \hline
    \end{tabular}
\caption{As in Table~\ref{table2}, but for the $z$-component of the fluid velocity shown in Fig.~\ref{figure10}.}\label{table3}
\end{table}

\begin{figure}[htp!]
	\begin{center}
	    {\includegraphics*[width=0.95\textwidth]{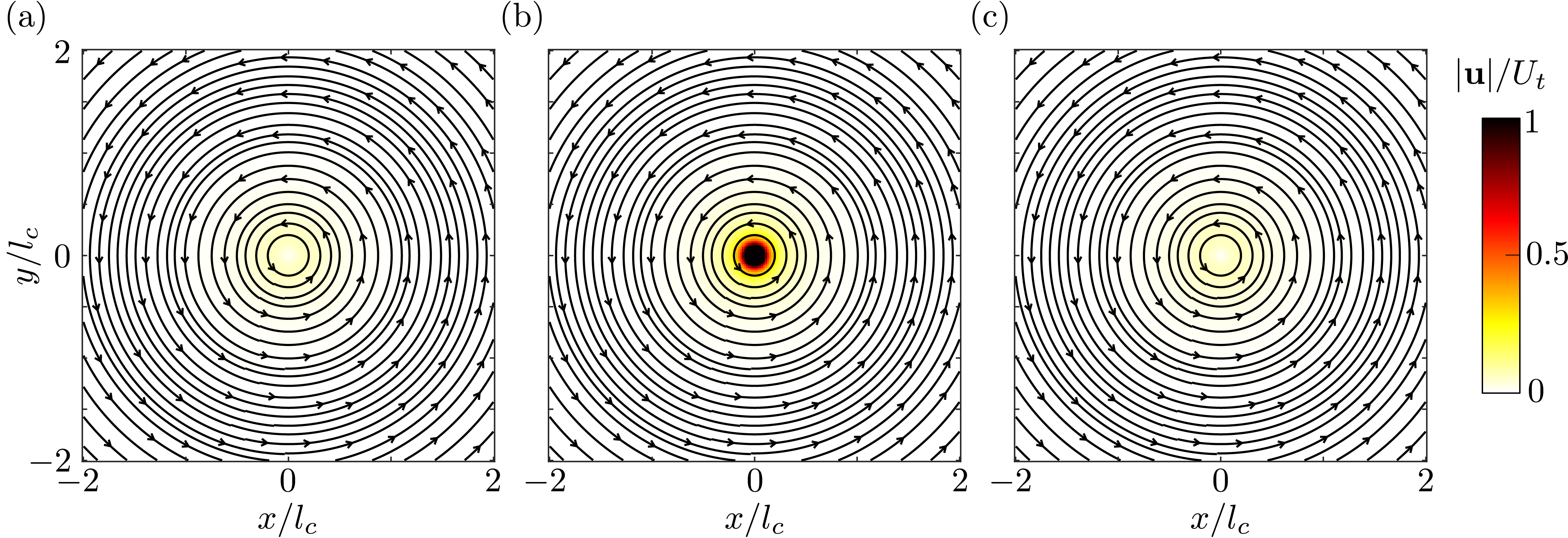}}
	\end{center}
	\caption{Heat map of the fluid velocity magnitude $|\bm{u}|/U_t$ and streamlines due to a point torque oriented perpendicular to walls ($\mathbf{\Omega}=|\mathbf{\Omega}|\hat{\mathbf{e}}_z$) separated by a distance of $H=1.26l_c$ at (a) $z=d-l_c/4$, (b) $z=d$ and (c) $z=d+l_c/4$ (where $d=0.629 l_c$). The remaining parameters are listed in Table~\ref{table1}.}\label{figure11}
\end{figure}

The fluid velocity is shown in Fig.~\ref{figure11} for the point torque oriented perpendicular to the rigid walls ($\mathbf{\Omega}=|\mathbf{\Omega}|\hat{\mathbf{e}}_z$) separated by distance $H=1.26l_c$. Results are shown for the three planes $z=d-l_c/4$, $z=d$, and $z=d+l_c/4$. The streamlines resemble those of an unbounded rotlet solution \cite{blake1974fundamental}, which decays as $1/r^2$, with the exception that its magnitude decreases when it approaches the walls. In contrast to Fig.~\ref{figure8}(e), there are no secondary vortices for the point torque oriented perpendicular to walls and positioned in the middle of the channel. Furthermore, no substantial changes in the direction of the flow occur in the vicinity of the no-slip walls.

\begin{figure}[!h]
	\begin{center}
	    {\includegraphics*[width=0.98\textwidth]{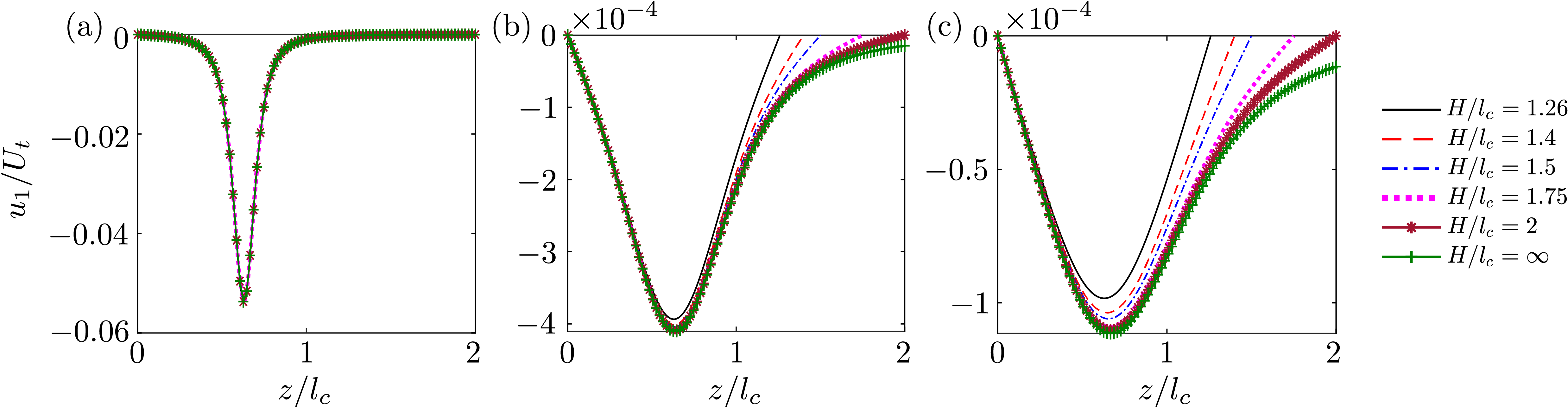}}
	\end{center}
	\caption{Variation of the scaled $x$-component of the velocity vector, i.e., $u_1/U_t$, with scaled $z/l_c$ in the $y=0.005l_c$ plane for various distances between parallel walls calculated at (a) $x=0.1l_c$, (b) $x=0.5l_c$ and (c) $x=0.75l_c$ from the position of the point torque at $x_r=y_r=0$ and $z_r=d/l_c$ (where $d=0.629 l_c$) oriented perpendicular to walls ($\mathbf{\Omega}=|\mathbf{\Omega}|\hat{\mathbf{e}}_z$). Model parameters are listed in Table~\ref{table1}.}\label{figure12}
\end{figure}

For the point torque aligned perpendicular to the rigid walls, we investigated how the components of the fluid velocity are influenced by confinement. The $x$-component of the velocity profile is shown as a function of $z$ in Fig.~\ref{figure12}, at three different $x$ positions throughout the channel (all with $y=0.005l_c$), and for various values of $H$. The profiles behave in a manner similar to those shown in Figs.~\ref{figure9} and \ref{figure10}, where the maximum speed, and coordinate at which it is attained, both increase with $H$. In addition, the velocity profile $u_1$ in the bounded domain approaches the profile in the semi-bounded domain ($H\to \infty$) for larger $H$.

To quantify the observations in Fig.~\ref{figure12}, we follow the same method as given in Eq.~\eqref{PD_for} for evaluating the percentage difference between the maxima of the velocity profiles $u_1$ associated with the semi-bounded and bounded domains, 
\begin{eqnarray}
   \text{PD}_x^{\perp}=\bigg|\frac{|u_1(\bm{x}_{\perp})|_{sb}-|u_1(\bm{\xi}_{\perp})|_b}{|u_1(\bm{x}_{\perp})|_{sb}}\bigg| \times 100\%  \label{PD_perp}
\end{eqnarray}
where $\bm{\xi}_{\perp}$ and $\bm{x}_{\perp}$ are the positions at which the magnitude of velocity components in the bounded domain and semi-bounded domain, respectively, attain the maximum values at fixed $x$ and $H$.\color{black}

\begin{table}[!h]
    \centering
    \begin{tabular}{|c|c|c|c|c|c|}
    \hline
        PD$_x^{\perp}$ & $H/l_c=1.26$ & $H/l_c=1.4$ & $H/l_c=1.5$ & $H/l_c=1.75$ & $H/l_c=2$\\ \hline
    $x/l_c=0.1$ & $0.46$ & $0.50$ & $0.49$ & $0.48$ & $0.002$\\ \hline
        $x/l_c=0.5$ & $4.10$ & $2.19$ & $1.46$ & $0.59$ & $0.31$\\ \hline
        $x/l_c=0.75$ & $11.80$ & $6.99$ & $4.92$ & $2.24$ & $1.10$\\
        \hline
    \end{tabular}
    \caption{Percentage difference (PD$_x^{\perp}$) between the maximum value of the fluid velocity $x$-component in the semi-bounded and the bounded domains found using Eq.~\eqref{PD_perp} for different $x$ and $H$. The corresponding velocity profiles are shown in Fig.~\ref{figure12}. Model parameters are given in Table~\ref{table1}.}\label{table4}
\end{table}

Table~\ref{table4} shows the percentage difference (PD$_x^{\perp}$) calculated for the components $u_1$ illustrated in Fig.~\ref{figure12}. The largest percentage difference observed is $11.80\%$ at $H=1.26l_c$ for $x=0.75l_c$. However, this difference is relatively small compared to the percentage difference observed for the point torque aligned parallel to the walls (Table~\ref{table2}). 

%---------------------------------------------------------------------------
\section{Conclusions}\label{S4}

In this paper, we have used various steady singularity solutions of the Stokes equations in order to capture the time-averaged flow generated by a cilium in both a semi-infinite domain, and a region confined between two plane parallel walls.

In Section~\ref{S2} we examined the flow fields associated with a point torque (rotlet), a single point force (Stokeslet), two point forces (two-Stokeslet), and four point forces (four-Stokeslet). This approach dramatically simplifies the time-varying and spatially-dependent distribution of forces exerted by a flagellum on the fluid, and is shown to accurately recreate the steady (i.e., time-averaged) flow across multiple length scales. Specifically, we assessed the minimum relative error between the singularity solutions for the velocity field, as compared to the widely studied colloidal rotor model. The rotlet and four-Stokeslet models provided excellent agreement, with optimal configurations yielding just 2.2\% and 1.9\% fitting error for the average velocity in the fluid domain where the optimization was performed, compared to the predictions of the colloidal rotor model. The single-Stokeslet and two-Stokeslet models incurred larger fitting errors of 8.8\% and 7.2\%, respectively. Importantly, however, the rotlet model had fewer fitting parameters (torque strength and position) compared to the four-Stokeslet model, because the latter required four point forces to be independently applied at distinct positions. Additionally, we have found that the streamlines predicted by the rotlet model closely resemble those due to {\it Volvox} flagella computed using the colloidal rotor model, at distances comparable to the length scale of individual flagella.

Notwithstanding the small temporal excursions due to the power and recovery strokes, the rotlet model also provides excellent predictions for experimental particle paths in the vicinity of the beating {\it Volvox} flagella \cite{pedley2016squirmers}. It has recently been shown that weakly inertial effects around beating {\it Chlamydomonas} flagella can give rise to small phase lags in the surrounding fluid and a spatial decay which is more rapid than in the zero Reynolds number flow \cite{wei2019zero}. However, since the time-averaged flow was shown to decay at the rate predicted in the limit of zero Reynolds number, our approach circumvents problems associated with inertia. These effects would likely come into play when considering the oscillatory flows relevant for synchronization, or the process of particle capture. Despite the substantial differences in power-recovery strokes and its trajectory (for example, arc \cite{katoh2018three} or circle \cite{brumley2014flagellar}), the time-averaged flow field decays at the same rate in both near- and far-field.

In Section~\ref{S3} we derived analytical expressions for the flow field generated by a point torque, situated between two plane parallel no-slip walls. This was accomplished using the method of images and Fourier transforms. Unlike previous papers on related topics \cite{hackborn_1990, van2007stokes, dauparas2016flagellar}, here we investigate the influence that plane, parallel no-slip boundaries have on the rotlet solution in 
both near- and far-field, for arbitrary orientation of the point torque. Predictably, the impact of the upper wall becomes less significant as the distance between the parallel walls increases and the flow field approaches that of a point torque located above a single no-slip wall. 
However, even when this distance is of the order of several ciliary lengths, we find that changes to the flow field caused by the upper wall are very small compared to the typical velocity scales of the ciliary tip.

The effect of hydrodynamic confinement was also shown to modify the structure of the resulting flow field. For a point torque oriented parallel to the walls, secondary vortices exist either directly above, or to the side of, the primary vortex, depending on the separation between the walls. While the existence of these closed streamlines would in principle have implications for particle transport within the flow, the consequences are likely to be minimal owing to the weak fluid velocity in those regions, and Brownian motion is likely to be more consequential. For a point torque oriented perpendicular to the walls, the counter-rotating secondary vortices do not exist, and the resulting flow is represented by circular streamlines.

The collective flows generated by arrays of cilia and flagella underpin a range of processes such as nutrient transport around corals \cite{ahmerkamp_simultaneous_2022} and mucociliary clearance in the respiratory tract \cite{juan2020multi}. In general, calculating the flow fields of such ciliary ensembles in a way that captures the near-field hydrodynamics can be computationally demanding \cite{Elgeti:2013}. The rotlet model presented in this paper provides an appealing solution for the efficient calculation of collective flows of large ciliary carpets. Because flows from individual cilia can be resolved, it would be straightforward to include heterogeneity in the spatial distribution and properties of cilia in these calculations. The expressions for the resulting fluid velocity can also be used to find the transport of dissolved organic matter, or passive or active particles undergoing advection and shear-reorientation \cite{Rusconi2014}. Thus, we expect our model to be broadly utilized in calculations of three-dimensional flows around ciliary carpets in both unbounded and confined geometries, and to be useful for investigating processes of transport, mixing, and feeding.

\begin{center}
\begin{Large}
{\sc Acknowledgments}
\end{Large}
\end{center}
The work was enabled by the dual-award Ph.D. program between the University of Manchester and the University of Melbourne. S.A.S. is grateful to the University of Melbourne for providing funding through its Xing Lei Scholarship and the Melbourne Research Scholarship. D.R.B. was supported by an Australian Research Council (ARC) Discovery Early Career Researcher Award No. DE180100911.

\newpage
\begin{center}
\begin{LARGE}
{\sc Supplementary Information}
\end{LARGE}
\end{center}

\setcounter{figure}{0}
\renewcommand{\thefigure}{S\arabic{figure}}
\setcounter{equation}{0}
\renewcommand{\theequation}{S\arabic{equation}}

\setcounter{section}{0}
\renewcommand{\thesection}{S\arabic{section}}

%---------------------------------------------------------------------------
\section{Numerical integration}\label{SecB}
In order to solve for the flow field given by Eq.~\eqref{ee17}, we need to integrate the expression Eq.~\eqref{e15} and Eq.~\eqref{ee28} for the singular $S_{ij}$ and the auxiliary $F_{ij}$ kernels, respectively.
However, the integrands in these equations become singular as variables $\lambda \to 0$ and $\zeta \to 0$. To avoid the singularity, we employ the asymptotic expansion to integrate the integrands in Eq.~\eqref{e15} and Eq.~\eqref{ee28} on intervals $\lambda \in [0,\epsilon_1]$ and $\zeta \in [0,\epsilon_2]$, respectively, where $\epsilon_\alpha \ll 1$, $\alpha \in \{1,2\}$. The remainders of Eq.~\eqref{e15} and Eq.~\eqref{ee28} are then calculated on intervals $\lambda \in [\epsilon_1,\lambda_{\infty})$ and $\zeta \in [\epsilon_1,\zeta_{\infty})$, respectively, for sufficiently large parameters $\lambda_{\infty}$ and $\zeta_{\infty}$, by employ the midpoint rule \cite{suli2003introduction}. 

The integrals for the singular $S_{ij}$ and auxiliary $F_{ij}$ kernels are rewritten as follow:
\begin{eqnarray}
    &&\int_0^{\infty}\lambda J_0(\tau \lambda) \frac{\sinh{\lambda (H-d)}}{\sinh{\lambda H}}\cosh{\lambda z} ~d\lambda=\frac{\epsilon_1^2}{2}\bigg(\frac{d}{H}-1\bigg) -\sum_{p=1}^{P_{\lambda}} \lambda^{(p)} J_0(\tau \lambda^{(p)}) \frac{\sinh{\lambda^{(p)} (H-d)}}{\sinh{\lambda^{(p)} H}}  \cosh{(\lambda^{(p)} z)} \Delta \lambda, \\
    &&\int_0^{\infty}\lambda J_0(\tau \lambda) \frac{\sinh{\lambda 
    d}}{\sinh{\lambda H}}\cosh{\lambda (H-z)} ~d\lambda=\frac{\epsilon_1^2}{2}\frac{d}{H} -\sum_{p=1}^{P_{\lambda}} \lambda^{(p)} J_0(\tau \lambda^{(p)}) \frac{\sinh{\lambda^{(p)} d}}{\sinh{\lambda^{(p)} H}} \cosh{(\lambda^{(p)} (H-z))} \Delta \lambda, \\
    &&\int_0^{\infty}\lambda J_1(\tau \lambda) \frac{\sinh{\lambda 
    d}}{\sinh{\lambda H}}\sinh{\lambda (H-z)} ~d\lambda= \sum_{p=1}^{P_{\lambda}} \lambda^{(p)} J_1(\tau \lambda^{(p)}) \frac{\sinh{\lambda^{(p)} (H-d)}}{\sinh{\lambda^{(p)} H}} \sinh{(\lambda^{(p)} z)} \Delta \lambda, \\
     && \int_0^{\infty}J_0(\tau \zeta) \phi(\zeta) ~d\zeta=\bigg(1-\frac{d}{H}-\frac{z}{H}\bigg) \frac{\epsilon_2^2}{2}+\sum_{p=1}^{P_{\zeta}}  J_0(\tau \zeta^{(p)}) \phi(\zeta^{(p)}) \Delta \zeta,\\
    &&\int_0^{\infty}J_1(\tau \zeta) \chi(\zeta) ~d\zeta=\sum_{p=1}^{P_{\zeta}}  J_1(\tau \zeta^{(p)}) \chi(\zeta^{(p)}) \Delta \zeta,\\
    &&\int_0^{\infty}J_1(\tau \zeta) \kappa(\zeta) ~d\zeta=\sum_{p=1}^{P_{\zeta}}  J_1(\tau \zeta^{(p)}) \kappa(\zeta^{(p)}) \Delta \zeta,\\
    &&\int_0^{\infty}J_1(\tau \zeta) \psi(\zeta) ~d\zeta=\sum_{p=1}^{P_{\zeta}}  J_1(\tau \zeta^{(p)}) \psi(\zeta^{(p)}) \Delta \zeta,
\end{eqnarray}
where $P_{\lambda}$ and $P_{\zeta}$ are the number of 
integration points in
$[\epsilon_1,~\lambda_{\infty}]$ and $[\epsilon_2,~\zeta_{\infty}]$, respectively, $\Delta \lambda  = (\lambda_{\infty}-\epsilon_1)/P_{\lambda}$ and $\Delta \zeta = (\zeta_{\infty}-\epsilon_2)/P_{\zeta}$ are the corresponding step sizes, $\lambda^{(p)}=\displaystyle \frac{1}{2}(\lambda^{(p-1)} +\lambda^{(p+1)})$ and $\zeta^{(p)}=\displaystyle \frac{1}{2}(\zeta^{(p-1)} +\zeta^{(p+1)})$ are the midpoints of $p$\textsuperscript{th} sub-intervals  $[\lambda^{(p-1)},~\lambda^{(p+1)}]$ and $[\zeta^{(p-1)},~\zeta^{(p+1)}]$, respectively. Throughout the paper, we have used $P_{\lambda}=P_{\zeta}=150$, $\epsilon_1=\epsilon_2=0.001 d$ and $\lambda_{\infty}=\zeta_{\infty}=300d$, and checked that the results are numerically converged by doubling the length of intervals and the number of integration points.

%---------------------------------------------------------------------------
\section{Deriving the solution for the flow in a semi-bounded domain}\label{A}

To validate the derivation in Section~\ref{S31}, we can compare the fluid velocity due to a point torque between two parallel walls obtained in Eq.~\eqref{ee17} to the existing theory on the fluid velocity due to a point torque above a single wall~\cite{blake1974fundamental}. This requires deriving the single wall approximation from the solution Eq.~\eqref{ee17}, by assuming that the distance between the two walls becomes infinite (i.e., $H \to \infty$). Taking $H \to \infty$ in Eq.~\eqref{ee17} modifies integrands in the singular kernel $S_{ij}$ (Eq.~\eqref{e15}) and the auxiliary kernel $F_{ij}$ (Eq.~\eqref{ee28}) as follows:
\begin{subequations}
\begin{eqnarray}
\lim_{H \to \infty}\bigg[\lambda J_0(\tau \lambda) \frac{\sinh{\lambda (H-d)}}{\sinh{\lambda H}} \cosh{\lambda z}\bigg]&=&\lambda J_0(\tau \lambda) \cosh{\lambda z}~ e^{-\lambda d},\label{F1a}\\
\lim_{H \to \infty}\bigg[\lambda J_0(\tau \lambda) \frac{\sinh{\lambda d}}{\sinh{\lambda H}} \cosh{\lambda (H-z)}\bigg]&=&\lambda J_0(\tau \lambda) \sinh{\lambda d}~ e^{-\lambda z},\label{F1b}\\
\lim_{H \to \infty}\bigg[\lambda J_1(\tau \lambda) \frac{\sinh{\lambda (H-d)}}{\sinh{\lambda H}} \sinh{\lambda z}\bigg]&=&\lambda J_1(\tau \lambda) \sinh{\lambda z}~ e^{-\lambda d},\label{F1c}\\
\lim_{H \to \infty}\bigg[\lambda J_1(\tau \lambda) \frac{\sinh{\lambda d}}{\sinh{\lambda H}} \sinh{\lambda (H-z)}\bigg]&=&\lambda J_1(\tau \lambda) \sinh{\lambda d}~ e^{-\lambda z},\label{F1d}\\
\lim_{H \to \infty} J_0(\tau \zeta) \phi(\zeta)&=&z \zeta J_0(\tau \zeta) e^{-\zeta(d+z)},\label{F1e}\\
\lim_{H \to \infty} J_1(\tau \zeta) \psi(\zeta)&=&z \zeta J_1(\tau \zeta) e^{-\zeta(d+z)},\label{F1f}
\\
\lim_{H \to \infty} J_1'(\tau \zeta) \psi(\zeta)&=&z \zeta J_1'(\tau \zeta) e^{-\zeta(d+z)},\label{F1g}\\
\lim_{H \to \infty} J_1(\tau \zeta) \kappa(\zeta)&=&z \zeta^2 J_1(\tau \zeta) e^{-\zeta(d+z)},   \label{F1h}
\end{eqnarray} \label{F1}
\end{subequations}
so that the following expressions for the singular kernel
\begin{subequations}
\begin{eqnarray}
S^{\infty}_{\alpha \beta}&=&\begin{cases}\displaystyle-\frac{\varepsilon_{\alpha \beta 3}}{4\pi \mu} \int_0^{\infty}\lambda J_0(\tau \lambda) \cosh{\lambda z}~ e^{-\lambda d}~d\lambda~~\mbox{for}~~z<d,
\\\vspace*{0.3cm}
\displaystyle \frac{\varepsilon_{\alpha \beta 3}}{4\pi \mu}\int_0^{\infty}\lambda J_0(\tau \lambda) \sinh{\lambda d}~ e^{-\lambda z} d\lambda~~\mbox{for}~~z\geq d,
\end{cases}\\\vspace{0.5cm}
S^{\infty}_{3 \alpha}&=&\begin{cases}\displaystyle \frac{\varepsilon_{3 \alpha \beta}}{4\pi \mu} \frac{r_{\beta}}{\tau} \int_0^{\infty}\lambda J_1(\tau \lambda) \sinh{\lambda z}~ e^{-\lambda d}~d\lambda~~\mbox{for}~~z<d,
\vspace{0.3cm}\\
\displaystyle \frac{\varepsilon_{3 \alpha \beta}}{4\pi \mu}  \frac{r_{\beta}}{\tau} \int_0^{\infty} \lambda J_1(\tau \lambda) \sinh{\lambda d}~ e^{-\lambda z} d\lambda~~\mbox{for}~~z\geq d,
\end{cases}\\\vspace{0.5cm}
S^{\infty}_{i3}&=&\begin{cases}\displaystyle \frac{\varepsilon_{i 3\alpha}}{4\pi \mu} \frac{r_{\alpha}}{\tau} \int_0^{\infty}\lambda J_1(\tau \lambda) \sinh{\lambda z}~ e^{-\lambda d}~d\lambda~~\mbox{for}~~z<d,
\vspace{0.3cm}\\
\displaystyle \frac{\varepsilon_{i 3\alpha}}{4\pi \mu}  \frac{r_{\alpha}}{\tau} \int_0^{\infty} \lambda J_1(\tau \lambda) \sinh{\lambda d}~ e^{-\lambda z} d\lambda~~\mbox{for}~~z\geq d,
\end{cases}
\end{eqnarray}\label{F2}
and the auxiliary kernel 
\end{subequations}
\begin{subequations}
\begin{eqnarray}
   F^{\infty}_{\alpha \beta}&=&\frac{\varepsilon_{k \beta 3}}{4\pi\mu} \bigg[\delta_{\alpha k} \int_0^{\infty}  \zeta J_0(\tau \zeta)~e^{-\zeta (d+z)}~d\zeta -\bigg[\frac{\delta_{\alpha k}}{\tau}
    -\frac{r_{\alpha} r_k}{\tau^3}\bigg]\int_0^{\infty} \zeta~z~J_1(\tau \zeta)~e^{-\zeta (d+z)}~d\zeta \nonumber\\
    &&-\frac{r_{\alpha} r_k}{\tau^2}\int_0^{\infty} \zeta z J_1'(\tau \zeta) e^{-\zeta (d+z)}~d\zeta \bigg], \\
    F^{\infty}_{3\alpha}&=& \frac{\varepsilon_{k\alpha 3}}{4\pi\mu}~\frac{r_k}{\tau}\int^{\infty}_0 \zeta^2 z J_1(\tau \zeta)~e^{-\zeta (d+z)}~d\zeta,~~~\mbox{and}~~~~F^{\infty}_{i3}=0
\end{eqnarray}\label{F3}
\end{subequations}
are obtained. 
Hence, the fluid velocity due to a point torque above the wall is \begin{eqnarray}
   u_i=(S^{\infty}_{ij}+F^{\infty}_{ij})\Omega_j.\label{F4}
\end{eqnarray}
This expression satisfies the appropriate boundary conditions on the wall (located at $z=0$) and at the infinite distance away from it:
$\bm{u}(x,y,0)=0$ and $\bm{u}(x,y,\infty)=0$.

\begin{figure}[htp!]
	\begin{center}
	    {\includegraphics*[width=0.5\textwidth]{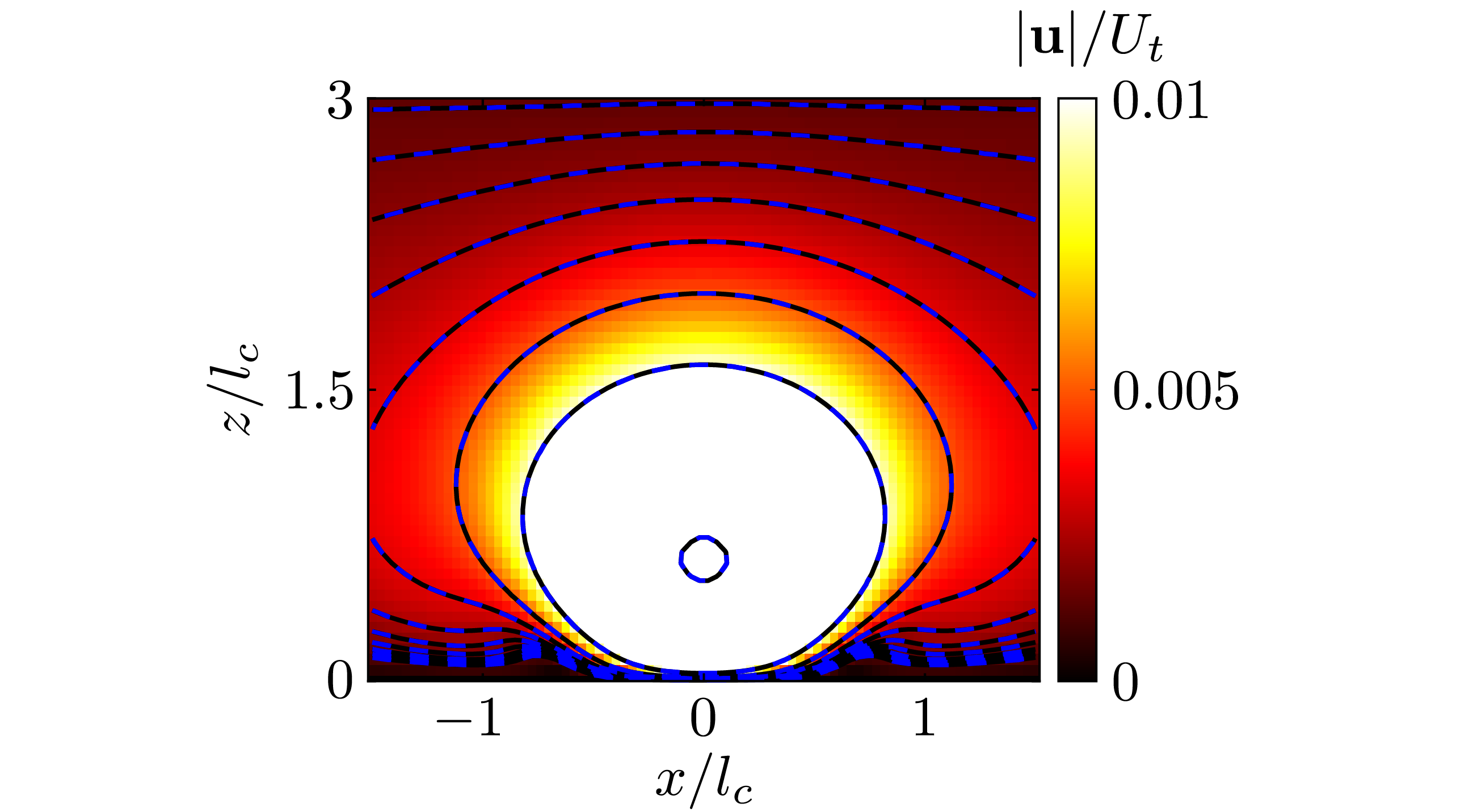}}
	\end{center}
	\caption{Heat map of the fluid velocity magnitude $|\bm{u}|/U_t$ due to the rotlet above the wall lying along $z=0$. The associated contours compare the velocity magnitudes obtained using Eq.~\eqref{F4}~[black solid line] and Eq.~\eqref{e5}~[blue dashed line] using the model parameters listed in Table~\ref{table1}. The displayed contour lines follow the sequence $\displaystyle |\bm{u}|/U_t:=\big\langle 1/(1+100(\ell-1)): \ell \in \mathbb{N} \big\rangle$, with velocity decreasing away from the point torque.}\label{FIGS1}
\end{figure}

The expression in Eq.~\eqref{F4} is compared to the classical result of \citet{blake1974fundamental} in Fig.~\ref{FIGS1}, where we plot the heat map of the fluid velocity magnitude and the associated contour lines, obtained using Eqs.~\eqref{F4} and \eqref{e5}, respectively, in the $y=0$ plane centered around the position of the point torque. The integration results are indistinguishable, suggesting that the flow field near a wall derived here agrees with the classical solution by \cite{blake1974fundamental}, for which the velocity field decays as $u\sim 1/r^2$ in both the near- and far-field. 

\newpage
%\bibliography{references}

%

\end{document}